\documentclass{SciPost}

\hypersetup{
    colorlinks,
    linkcolor={red!50!black},
    citecolor={blue!50!black},
    urlcolor={blue!80!black}
}

\usepackage[bitstream-charter]{mathdesign}
\usepackage{braket}
\usepackage{color}
\usepackage{latexsym}
\usepackage{graphicx}
\usepackage{float}
\usepackage{dcolumn}
\usepackage{bm}

\usepackage{amsmath}
\usepackage{mathtools}
\usepackage[space]{grffile}
\usepackage{xcolor}

\usepackage{comment}
\usepackage{physics}
\usepackage[colorinlistoftodos]{todonotes}

\def\XXint#1#2#3{{\setbox0=\hbox{$#1{#2#3}{\int}$}
     \vcenter{\hbox{$#2#3$}}\kern-.5\wd0}}

\newcommand{\niceref}[1] {Eq.~(\ref{#1})}

\newcommand{\be}{\begin{equation}}
\newcommand{\ee}{\end{equation}}
\newcommand{\bea}{\begin{align}}
\newcommand{\eea}{\end{align}}

\newcommand{\phd}{{\phantom{\dagger}}}

\newcommand{\ii}{\mathrm{i}}

\fancypagestyle{SPstyle}{
\fancyhf{}
\lhead{\colorbox{scipostblue}{\bf \color{white} ~SciPost Physics }}
\rhead{{\bf \color{scipostdeepblue} ~Submission }}

\fancyfoot[C]{\textbf{\thepage}}
}

\begin{document}

\pagestyle{SPstyle}

\begin{center}{\Large \textbf{\color{scipostdeepblue}{
Logarithmic singularity in a dynamical quantum phase transition for free fermions}}}\end{center}

\begin{center}\textbf{
Y. Bezzaz\textsuperscript{1*},
D. M. Gangardt\textsuperscript{1},
P. L. Krapivsky\textsuperscript{2},
J. M. Luck\textsuperscript{3},
K. Mallick\textsuperscript{3},
S. Prolhac\textsuperscript{4}
}\end{center}

\begin{center}
{\bf 1} School of Physics and Astronomy, University of Birmingham, Edgbaston,
Birmingham, B15 2TT, UK
\\
{\bf 2} Santa Fe Institute, Santa Fe, New Mexico 87501, USA
\\
{\bf 3} Institut de Physique Théorique, Université Paris-Saclay, CEA and CNRS,
91191 Gif-sur-Yvette, France
\\
{\bf 4} Laboratoire  de Physique Théorique, Université de Toulouse, 31062 Toulouse Cedex 04, France

\bigskip
$\star$ \href{mailto:yxb207@student.bham.ac.uk}{\small yxb207@student.bham.ac.uk}
\end{center}

\section*{\color{scipostdeepblue}{Abstract}}

\textbf{%
We study the Loschmidt echo in a system of $N$ non-interacting spinless lattice fermions released from a double-domain-wall initial state. In the large $N$ limit, 
the return probability is characterized  by a large-deviation rate function known as the dynamical free energy. The Loschmidt echo is dominated by a complex instanton configuration, and the dynamical free energy develops a logarithmic singularity at the dynamical quantum phase transition. We obtain analytical results in the short-time and long-time regimes and support them with numerical computations.  We show that the transition can be understood as a topological change of the dominant instanton configuration, analogous to the emergence of a cut in random matrix theory. We further show that post-selection can drive the system through two successive DQPTs, associated with successive topological changes of the dominant instanton configuration in complex time.}

\vspace{\baselineskip}

\vspace{10pt}
\noindent\rule{\textwidth}{1pt}
\tableofcontents
\noindent\rule{\textwidth}{1pt}
\vspace{10pt}

\section{Introduction}

Understanding the dynamics of many-body quantum systems is central in modern theoretical physics. Such systems give rise to a wealth of collective phenomena in condensed matter, quantum liquids, and ultracold atomic gases, yet their theoretical description is notoriously difficult due to the exponential growth of the Hilbert space and the crucial role of quantum entanglement. Recent experimental advances, particularly in ultracold atom platforms, together with progress in numerical and analytical techniques, have enabled detailed studies of nonequilibrium dynamics, thermalization, and information spreading in isolated quantum systems \cite{Polkovnikov2011,Eisert2015,Bloch2012}. These developments have stimulated renewed efforts to identify unifying principles governing the emergence of effective classical behavior from microscopic quantum laws, with important implications for both fundamental physics and the development of quantum technologies \cite{Gross2017,Preskill2018,Altman2021}.

Quantum quenches provide a natural and clean setting for studying these questions. In a quench protocol, the system is prepared in an initial state $\ket{\Psi_0}$, typically the ground state of an initial Hamiltonian $H_0$. At time $t=0$, the Hamiltonian is suddenly changed to $H$, and the state evolves unitarily as
$$
\ket{\Psi(t)} = e^{-\ii Ht}\ket{\Psi_0}\, .
$$
Nonequilibrium quantum many-body dynamics following a quench may undergo a dynamical quantum phase transition (DQPT) manifested by non-analyticity of a dynamical observable at a certain time $t_c$ \cite{Heyl2013,Heyl2018}. A natural diagnostic for detecting DQPTs is the return probability to the initial state, known as the \emph{Loschmidt echo}, which is defined as 
\begin{equation}
L_N(t)=|A_N(t)|^2= \left|\bra{\Psi_0}e^{-\ii Ht}\ket{\Psi_0}\right|^2 \, ,
\end{equation}
where $A_N(t)$ denotes the \emph{Loschmidt amplitude}, i.e., the transition amplitude from the state $\ket{\Psi_0}$ back to itself after a time $t$, and $N$ represents the number of degrees of freedom of the system. Unlike local observables, the Loschmidt echo probes global properties of the evolving quantum state and encodes highly nontrivial many-body dynamical correlations. Typically, the Loschmidt echo takes a form 
\begin{equation}
  L_N(t) \simeq \exp\!\left[-\,2 N^{\alpha}\, S\!\left(t/N^{\beta}\right)\right],
  \label{LDAmplitude}
\end{equation}
in the scaling limit
\begin{equation}
\label{scaling}
t\to\infty, \quad N\to\infty, \quad t/N^{\beta}=\text{finite} \, .
\end{equation}
The exponents $\alpha$ and $\beta$ depend on the model and on the choice of initial state.  Because $L_N(t)$ is exponentially small at large $N$, it naturally corresponds to a rare-event probability. This draws a clear connection with large-deviation theory, where rare dynamical events are governed by an extensive rate function \cite{Touchette2009}. In the quantum context,  time plays the role of inverse temperature and the function $S$ is a rate function;  within this perspective, a DQPT corresponds to a non-analyticity of $S$ (or its derivatives) at a critical time $t_c$. Quantum dynamical transitions have by now been identified in a broad range of models, including quantum spin chains, topological systems, and driven many-body systems \cite{Heyl2013,Heyl2018,Parez2022,Parez2026}. Their observation is challenging in many-body settings, yet recent advances in quantum-simulation platforms, most notably quantum gas microscopes and trapped-ion systems, have enabled measurements of Loschmidt echoes and the experimental observation of DQPTs \cite{jurcevic2017,flaschner_observation_2018, guoObservationDynamicalQuantum2019}.

Motivated by these developments, we investigate the emergence of dynamical singularities in a simple, analytically tractable system of non-interacting fermions. We consider the evolution of $N$  spinless fermions on a one-dimensional lattice, starting from the initial double-domain-wall (DDW) configuration, i.e., with $N$ adjacent sites initially occupied. The evolution of this configuration was investigated in \cite{KLM18}, where it was shown that in the $t\to\infty$ limit, the return probability decays algebraically, $L_N\propto t^{-\gamma_N}$. The decay exponent $\gamma_N=\left\lfloor \frac{N^2 + 1}{2}  \right\rfloor$ depends on the parity of the fermion number. Furthermore, the decay is algebraic only for even $N$; for odd $N$, the power law decay is modulated by periodic oscillations.  The scaling behavior \eqref{LDAmplitude} is characterized by the exponents $(\alpha, \beta)=(2, 1)$. This was mentioned in \cite{KLM18}, where the asymptotic behaviors of the rate functions were also probed. 

A recent study~\cite{perez-garcia} identified a DQPT at a critical rescaled time  $\tau_c$, with $\tau=t/N$, through an analytic continuation of the Gross-Witten-Wadia model\cite{GrossWitten1980}.
However, the dominant saddle-point configuration in the long-time regime, and hence the nature of the real-time non-analyticity, remained undetermined. In the present work we determine  the dominant
saddle-point configuration  on both sides of the transition and show that the DQPT originates from a change of topology
of  the instanton configuration, analogous to the
\emph{birth of a cut} in random matrix theory. Due to the  distinctive  shapes  of the corresponding Coulomb gas configurations, we refer to the short-time and  the long-time regimes as "unbroken-heart" and "broken-heart" regimes respectively: in the former, the instanton support forms a single heart-shaped  contour, while in the latter the contour  splits into two components.
The topological transition between these regimes gives  rise to a
logarithmic singularity of the dynamical free energy near $\tau_c$, leading to
a phase transition of order $2+0$.  Logarithmic singularities were reported  in several QPTs, e.g., in models of quantum spin glasses \cite{Huse_SG,Read_SG}. 

We further extend the analysis to complex time,  motivated by the interpretation of the complex-time Loschmidt amplitude in terms of a post-selected work distribution. There the  resulting phase diagram exhibits the  complex-coupling GWW phase structure identified previously
in Ref.~\cite{copetti2022delayed}. In this  setting, we explicitly follow the evolution of the relevant saddle-point configuration 
 across the different phases and show that the second
phase boundary corresponds to a \emph{death of a cut} transition. This transition is  accompanied by the
same logarithmic singularity and is therefore also of order $2+0$.

The outline of this work is as follows. In Section~\ref{sec:setup}, we define the model,
review previous results, and summarize our main findings. In Section~\ref{sec:Coulomb},
we derive the large-$N$ behavior of the Loschmidt amplitude by mapping it to a Coulomb
gas problem and explicitly determining the dominant complex instanton configuration. In
Section~\ref{sec:DQPT}, we analyze the real-time DQPT and show that it corresponds to
a topological change of the  instanton configuration.   We identify this transition with the \emph{birth of a cut} phenomenon
and derive the associated logarithmic singularity of the dynamical
free energy. In
Section~\ref{sec:complex_time} we extend the analysis
to complex time. We study the resulting phase structure, track the evolution of the dominant
saddle across its different topologies, and characterize the second \emph{death of a cut}
transition. Finally, in Section~\ref{sec:disc}, we summarize our results and discuss possible
extensions. Several technical derivations are relegated to the Appendices.

\section{Loschmidt echo in a free fermionic chain}
\label{sec:setup}

We consider a system of \( N \) non-interacting spinless fermions on a one-dimensional infinite lattice subject to the tight-binding Hamiltonian
\begin{equation}
\label{eq:ham}
    H =   \sum_{x\in \mathbb{Z}} \left( c_x^\dagger c^\phd_{x+1} + c_{x+1}^\dagger c^\phd_x \right) 
    = \int_{-\pi}^{\pi} \frac{dq}{2\pi}\, \varepsilon(q)\, c^\dagger(q) c(q)  \, , 
\end{equation}
with dispersion relation $ \varepsilon(q) = 2\cos q$. The fermionic operators $c_x$ and their Fourier transform  $c(q) = \sum_{x\in \mathbb{Z}} e^{-\ii q x} c_x $ obey  the canonical anti-commutation relations $ \qty{ c^\phd_x, c_y^\dagger } = \delta_{xy}$ and  
$\qty{c(q), c^\dagger(q') } = 2\pi \delta(q - q')$.
Besides, we have $ \qty{ c^\phd_x, c^\phd_y } = \qty{ c_x^\dagger, c_y^\dagger } = 0 .$

\begin{figure}[h]
    \centering
    \includegraphics[width=0.65\textwidth]{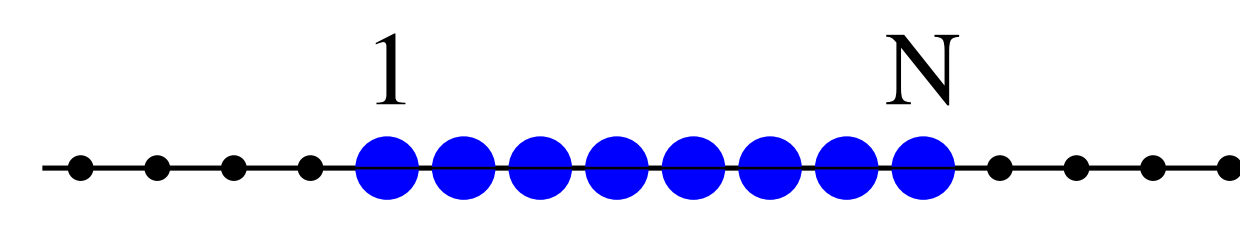}
    \caption{Initial DDW configuration of the $N$ fermions on an infinite one-dimensional lattice with sites $x=1,\ldots,N$ occupied, while all other sites are empty.}
    \label{fig:DW}
\end{figure}

We are interested in the return probability of the set of $N$ fermions to the initial state under the free evolution \eqref{eq:ham}. We consider the maximally dense 
initial state (Fig.~\ref{fig:DW}), viz., a DDW initial condition 
\begin{align}
\label{IC}
\ket{\Psi_0}=\prod_{x=1}^{N} c^\dagger_x \ket{0}    \, , 
\end{align}
with $N$ fermions initially occupying an interval of $N$ consecutive sites. 
The state \eqref{IC} can be viewed as a real-space analog of a Fermi sea, Fig.~\ref{fig:DW}.

There are only two parameters in our setting, the total number of particles $N$ and time $t$. We are interested in the situation when these parameters are both large, $N\to\infty$ and $t\to\infty$. We will see that in this situation, the behavior depends on the rescaled time $\tau = t/N$, corresponding to the critical exponent  $\beta=1$ reflecting the ballistic nature of our nearest-neighbor hopping dynamics.

\subsection{Summary of known results}

It was shown in Ref.~\cite{viti2016} that for a single domain-wall (DW) initial state in which all sites with  $x \ge 0$ are occupied and all sites with $x <0$ are empty, the Loschmidt echo exhibits a Gaussian decay
\begin{align}
\label{eq:1dw}
    L(t)=e^{-t^2} \,, 
\end{align}
which is exact for all $t\geq 0$. Equation \eqref{eq:1dw} can be reinterpreted \cite{Allegra2016} as a Wick-rotated version of the partition function of the XX chain with DW boundary conditions in imaginary time. In this framework, the model exhibits a remarkable limit-shape phenomenon \cite{stephan_extreme_2021}, in which the partition function is dominated by a single macroscopic configuration with sharp boundaries 
separating frozen and fluctuating regions.  

More recently, the evolution of the DDW initial configuration, Fig.~\ref{fig:DW}, was investigated in \cite{KLM18} in the limit $t \to \infty$ while keeping the fermion number $N$ finite. The calculations relied on the celebrated Selberg-Mehta integrals and showed that the Loschmidt echo decays as a power law when $t\to\infty$, with an exponent depending on the parity of $N$, and with an amplitude $K_N$ expressed in terms of the Barnes \(G\)-function.
\begin{align}
  L_N(t) \simeq \frac{K_N}{t^{\gamma_N}} \, ,\quad \gamma_N  =  \left\lfloor \frac{N^2 + 1}{2}  \right\rfloor,
  \label{eq:LE_long} 
\end{align} 
In the complementary limit of $N \to \infty$, with $t$ kept finite, we have  
\begin{align}
\label{eq:LE_short}
\lim_{N\to\infty} L_N(t)=e^{-2t^2} \ .
\end{align}
Indeed, if $t$ is finite, a very large DDW behaves as two independent single walls that contribute independently to the Loschmidt echo, so \eqref{eq:LE_short} is just the square of  \eqref{eq:1dw}. A formal proof can be obtained, e.g., by a straightforward extension of the methods of Ref.~\cite{viti2016}, based on the strong Szeg\"o theorem governing the asymptotic behavior of large Toeplitz determinants, see Appendix~\ref{Appendix : Scego} for details. 

The asymptotic formulas \eqref{eq:LE_long} and \eqref{eq:LE_short}  imply that the limits $t \to \infty$  and  $N \to \infty$  do not commute. However, the two limiting cases can be reconciled by writing  a large-deviation expression of the form \eqref{LDAmplitude}, namely
\begin{align}
\label{LN:scaling}
   L_N(t) \simeq \exp\!\left[- 2 \,N^2\, S\!\left(\tau\right)\right],
\end{align}
in the scaling limit \eqref{scaling} with $\beta=1$, i.e., with ratio $\tau=t/N$ kept finite. By comparing this scaling form with \niceref{eq:LE_long}  and \niceref{eq:LE_short}, one deduces \cite{KLM18} the asymptotic behaviors $S(\tau) \simeq \tau^2$ when $\tau \ll 1$ and 
$S(\tau) \simeq \frac{1}{4} \log\tau + \frac{3}{8}$ when $\tau \gg 1$. The full calculation of the  function $S$  and the existence of a critical value  $\tau_c$ where $S$ would be non-analytic are the main  questions we address in the present work.
 
It is  tempting to compare these results  to the behavior found in imaginary-time evolution of the free-fermionic XX chain with the DDW boundary conditions. This problem was recently considered in Ref.~\cite{pallister_limit_2022}, where a third-order phase transition was discovered by an exact mapping to the Gross-Witten-Wadia (GWW)  model of lattice QCD~\cite{GrossWitten1980,WADIA}. This transition was interpreted geometrically as the merger of two limit shapes emerging from the two domain walls in the boundary conditions. Hence, the correspondence between imaginary and real-time evolution suggested in Ref.~\cite{Allegra2016,pallister_limit_2022} via a Wick rotation hints at  a possible phase transition (DQPT) between the two regimes of the Loschmidt echo at some critical value $\tau_c$. Indeed, such a transition was reported in Ref.~\cite{perez-garcia} by using the fact that 
the  Loschmidt amplitude is the analytic continuation of  the Gross-Witten-Wadia (GWW) model to a purely imaginary coupling (identified with the real-time quantum evolution). This analytic continuation allows one to reproduce the short-time Gaussian behavior \eqref{eq:LE_short} and predicts a third-order transition for $\tau_{\mathrm{c}}\approx0.331$. However, the long-time behavior and, more generally, the full expression
of $S$ remained out of reach.

\subsection{Main new results}
\subsubsection{DQPT as a topological change of the instanton configuration}

By mapping the Loschmidt amplitude to a Coulomb gas model 
in the complex plane $z=e^{iq}$, 
we show below that in the large-$N$ limit, the Loschmidt echo is dominated  by an optimal distribution of $N$ charges and each charge represents a particle moving ballistically with momentum $q$. 
Since the Loschmidt echo  corresponds to a very atypical process, we find that the optimal configuration  requires the momenta $q$ to acquire a nonzero imaginary part pushing  $z=e^{iq} $ off the unit circle. This is in contrast with typical configurations in the free expansion, where only the initial state is fixed, so that the dominant momenta remain real and the corresponding charges stay on the unit circle.

\begin{figure}[h]
    \centering
    \includegraphics[width=1\textwidth]{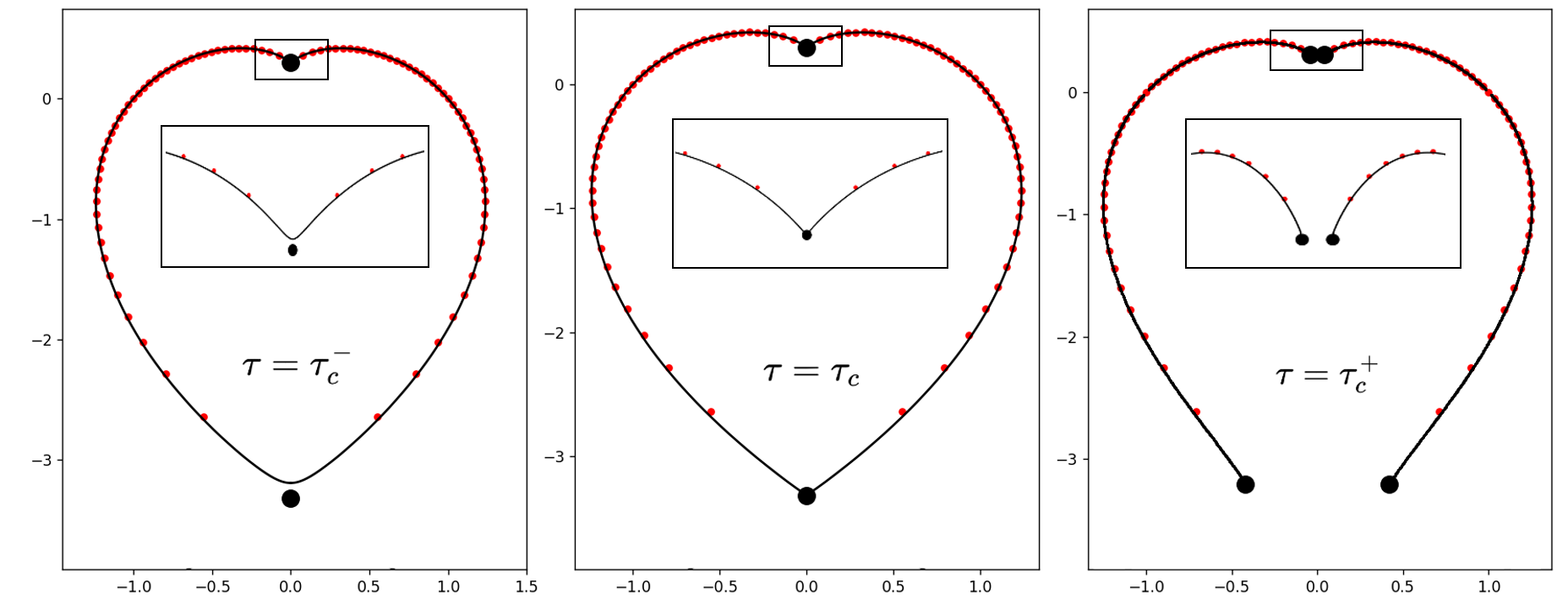}
\caption{Topological change of the dominant
instanton configuration across the critical rescaled time
$\tau=\tau_c$ in the complex plane $z=e^{iq}$. Red dots show the discrete  saddle points from solving numerically \eqref{eq:saddle point q}  for
$N=90$. The curves in black represent the density support of the saddle points in the continuous limit by solving \eqref{Anti stockes line}.}
\label{fig:broken_heart}
\end{figure}

In this picture,  the DQPT  reported in Ref.~\cite{perez-garcia} corresponds to a topological change of the dominant instanton configuration as illustrated in Fig.~\ref{fig:broken_heart}. The critical value $\tau_c \approx 0.331$ separates the dynamics into a short-time regime ($\tau<\tau_c$) and a long-time regime ($\tau>\tau_c$). The short-time regime $(\tau<\tau_c)$ corresponds to a dominant  one-cut configuration with a complex ungapped support. As $\tau$ increases, this support deforms continuously until it takes the shape of a heart at $\tau=\tau_c$. The Gaussian decay result \niceref{eq:LE_short} holds in the range $\tau \leq \tau_c$. In the long-time regime $(\tau>\tau_c)$, the Loschmidt echo is dominated by a two-cut instanton configuration, and the expression of $S$ is more involved as we show below.  At $\tau=\tau_c^{+}$, the support splits into two disconnected symmetric cuts, with $N/2$ charges on each cut, resembling a broken-heart. In what follows we  denote the short-time and the long-time regimes as the \emph{unbroken-heart} and  \emph{broken-heart} phases respectively. 

This topological change of the dominant instanton configuration illustrated in  Fig.~\ref{fig:broken_heart} is the mechanism behind the DQPT. We emphasize that this  transition is topologically distinct from the one arising in imaginary time \cite{pallister_limit_2022}. In the latter case, opening of a gap in the charge support does not affect the  genus of the corresponding Riemann surface. Here, the transition corresponds to  the \emph{birth of a cut} phenomenon known from random matrix theory \cite{eynardUniversalDistributionRandom2006,claeysBirthCutUnitary2008} in which the genus increases by one. We show below that this leads to a logarithmic correction near the critical point $\tau_c$:
\begin{equation}
\label{eq:derive action}
\frac{d S}{d \tau}=
\begin{cases}
2\tau & \tau \to \tau_c^{-} \\[6pt]
2\tau+\dfrac{ 2 (1+4 \tau_c^2) }{ \tau_c^2 }\dfrac{(\tau-\tau_c)}{\log(\tau-\tau_c)} & \tau \to \tau_c^{+} \, , 
\end{cases}
\end{equation}
 The second derivative of the singular correction vanishes as $1/\log(\tau-\tau_c)$, i.e., slower than any power of $(\tau-\tau_c)$. Therefore, the phase transition is infinitesimally weaker than standard second-order transition, justifying the term "phase transition of (2+0)-order". Thus  this phase transition is more singular than the third-order GWW phase transition \cite{GrossWitten1980,WADIA}.

\subsubsection{Complex-time phase structure and post-selected work statistics}
\label{section complex times post selection}

The above results indicate that the Loschmidt echo behaves qualitatively differently in real and imaginary time,  and that the two cases cannot be related by a naive analytic continuation (i.e Wick rotation). This motivates  us to extend the analysis to complex times $t\in\mathbb{C}$. Such an extension  also has a direct physical interpretation, since the Loschmidt amplitude at complex time can be related to  a post-selected work distribution \cite{Heyl2013}. Indeed, for a sudden quench from an initial state $\ket{\Psi_0}$, the characteristic function of work is given by the Loschmidt amplitude \cite{silva08} :
\begin{equation}
\chi(t) =\int dW\, e^{\ii Wt} P(W)=\bra{\Psi_0}e^{-\ii Ht}\ket{\Psi_0} \equiv A_N(t) \, .
\label{travail}
\end{equation}
A convenient way to bias the work statistics is to introduce  the
tilted distribution
\begin{equation}
    \widetilde{P}_{\beta}(W)
    =
    \frac{1}{A_N(i\beta)}\,
    P(W)e^{-\beta W} \, .
\end{equation}
The parameter \(\beta\) controls the bias towards different sectors of the work distribution and thus provides a natural implementation of post-selection. The normalized characteristic function associated with \(\widetilde P_\beta(W)\) is
\begin{equation}
    \chi_\beta(t)=\frac{1}{A_N(i\beta)}\int dW\, P(W)e^{\ii W(t+\ii \beta)}
    =\frac{A_N(t+i\beta)}{A_N(i\beta)} \,.
\end{equation}
Thus, post-selected work statistics is naturally encoded in the complex time 
Loschmidt amplitude.  

\begin{figure}[h]
    \centering
\includegraphics[width=0.7\textwidth]{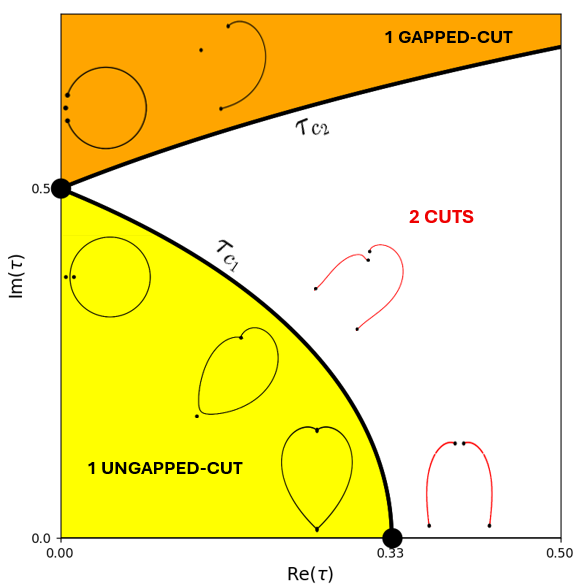}
\caption{Phase diagram of the dominant instanton configuration for  $A_N(\tau)$ in the complex plane of $\tau$. Typical shapes of the charge support are shown in each region. The point of merger of the two critical lines for  $\tau=i/2$ corresponds to the third-order Gross-Witten-Wadia transition. }
    \label{fig:phase_diagram}
\end{figure}
We obtain the phase diagram of the Loschmidt amplitude $A_N(\tau)$
as a function of the complex rescaled time\footnote{In the following, whenever \(\tau\) and \(\phi\) are used as separate variables, we denote the modulus \(|\tau|\) simply by \(\tau\).} 
$\tau=t/N=|\tau|e^{i\phi}$, shown in Fig.~\ref{fig:phase_diagram}.
In particular, the phase diagram shows how the topology of the dominant instanton configuration evolves throughout the complex-time plane. Three distinct phases emerge. The yellow region corresponds to the short-time regime, governed by a one-cut ungapped configuration. At the first critical line $\tau_{c_1}$, the support splits into two disconnected components, giving rise to an intermediate two-cut phase. Then at the second critical line $\tau_{c_2}$, one of the two cuts becomes completely depleted and the support reduces to a one-cut gapped configuration, shown in orange. The two critical lines therefore correspond respectively to a birth/death of a cut transition. Thus, post-selection can drive the evolution across both critical boundaries, leading to two successive DQPTs of order $2+0$. 

On the purely imaginary time line, the two critical lines merge and the transition reduces to the third-order GWW transition, for which the logarithmic corrections disappear.

The phase diagram also clarifies the limitations of a naive Wick rotation. While it correctly captures the one-cut phases, it fails in the intermediate two-cut regime, where the dominant configuration requires a redistribution of charges between the two components of the support.

\section{From the Loschmidt echo to the Coulomb gas}
\label{sec:Coulomb}
  
In contrast to approaches based on analytic continuation from imaginary time \cite{perez-garcia}, we derive the large-$N$ behavior of the Loschmidt echo by mapping it to a Coulomb gas problem and explicitly determining the dominant complex instanton configuration.

\subsection{Integral representation of the Loschmidt amplitude}
The central quantities of interest are the Loschmidt amplitude and the corresponding Loschmidt echo
\begin{equation}
A_N(t)
=
\expval{e^{-\ii tH}}{\Psi_0},
\qquad
L_N(t)=\qty|A_N(t)|^2.
\label{Loschmidt-def}
\end{equation}

For the state $\ket{\Psi_0}$ given by Eq.~\eqref{IC} one can use Wick's theorem to express  the Loschmidt amplitude \( A_N(t) \)  as an \(N\times N\) Toeplitz determinant of the Bessel functions of the first kind:
\begin{align}
    \label{eq:toeplitz}
    A_N(t) =
\det\!\left[\langle 0|\, c_m(t)\, c_n^\dagger(0)\, |0\rangle \right]_{1\le m,n\le N} =\det\big[J_{m-n}(2t)\big]_{1\le m,n\le N} \, . 
\end{align}
While we do not rely on the Toeplitz determinant representation in the main body of this work, we outline in Appendix~\ref{Appendix:Toeplitz} how the strong Szeg\"o theorem allows one to deduce the Gaussian short-time phase and how the critical value $\tau_c$ can be inferred from an exact recursion relation satisfied by the Toeplitz determinant~\niceref{eq:toeplitz}.
Instead we write the initial state in the $N$-particle momentum basis and  obtain 
\begin{equation}
\label{eq:LA}
A_N(t)=\int\frac{d{\bf q}}{(2\pi)^N}\,
\exp\!\left(-2\ii t\sum_{n=1}^N\cos q_n\right)\,
\big|\Psi_0({\bf q})\big|^2 ,
\end{equation}
where ${\bf q}=\{q_1,\ldots,q_N\}$ are the momenta of the non-interacting fermions and we employ the shorthand notation 
$\int d{\bf q} = \int_{-\pi}^{\pi}dq_1\cdots \int_{-\pi}^{\pi}dq_N$. 
  The normalized antisymmetric wavefunction $\Psi_0({\bf q})$ of $N$ fermions  takes the form of a
Vandermonde determinant:
\begin{align}
\label{eq:vandermond}
\Psi_0({\bf q}) = \frac{1}{\sqrt{N!}}\det \qty[e^{\ii m q_n }]_{m,n=1}^{N} =  \frac{1}{\sqrt{N!}} \, \prod_{n=1}^{N} e^{\ii q_n} 
\prod_{1\le m<n\le N}\qty(e^{\ii q_m}-e^{\ii q_n }) \ .
\end{align}
For large times, the phase in \eqref{eq:LA}  oscillates wildly, so that the integrals are dominated by stationary configurations corresponding to ballistic classical trajectories of the particles. So far this is nothing but the standard WKB approach to quantum mechanics in $N$-dimensional space. However, one must be careful in the large-$N$ limit: the Vandermonde determinant contribution  cannot be treated as a mere prefactor anymore, as it is a product of $N(N-1)/2=O(N^2)$ terms. Instead, it must be fully taken into account in determining the stationary configuration. 

Accordingly, we consider the double-scaling limit $(N,t)\to\infty$ with $\tau=t/N$ fixed. In this limit, it is convenient to rewrite the amplitude as
\begin{equation}
\label{eq:LA_rearr}
A_N(t)=\frac{1}{N!}\int\frac{d{\bf q}}{(2\pi)^N}\, e^{-N^2 S[{\bf q}]}\, \ ,
\end{equation}
where $S[{\bf q}]$ is the action of order $O(1)$:
\begin{equation}
\label{eq:LA_act}
        S[{\bf q}] = \frac{2\ii \tau}{N} \sum_{n=1}^{N}  \cos q_n-\frac{1}{N^2}\sum_{1\le m<n\le N}\log \qty|e^{\ii q_m}-e^{\ii q_n }|^2  \ .
\end{equation}
For a fixed value of $\tau$, the two terms in the action compete: the first tends to confine the charges near its saddle at points $q=0$ and $q=\pi$, whereas
the second one favors their spreading to reduce the repulsive logarithmic interactions.

The dynamical free energy $S(\tau)$ is determined by the dominant saddle-point configuration
\begin{equation}
    S(\tau)=S[{\bf q}^*] \ ,
\end{equation}
with the $q_n^*$ solution of
\begin{equation}
\label{eq:saddle point q}
    \frac{\partial S}{\partial q_n}[{\bf q}^*]=0\, \qquad n=1,\ldots,N\,.
\end{equation}
Thus, $S(\tau)$ depends on $\tau$ both explicitly from the factor in front of the first sum in (\ref{eq:LA_act}), and implicitly through the optimal $q_n^*(\tau)$ solution of (\ref{eq:saddle point q}). However its total derivative has a particularly simple form due to the stationarity condition \eqref{eq:saddle point q} :
\begin{equation}
\label{eq:S'}
    S'(\tau)= \frac{2\ii}{N} \sum_{n=1}^{N}\cos q_n^*\, .
    \end{equation}

\subsection{Large-N limit: the Coulomb gas}

As noted in Refs.~\cite{KLM18,perez-garcia}, the multiple-integral representation \eqref{eq:LA} of the Loschmidt amplitude has the structure of a Coulomb gas describing $N$ charges $z_n=\mathrm{e}^{\ii q_n}$ on the unit circle, subject to a complex external potential $V(q)=2\ii t\cos q$ and interacting through a repulsive logarithmic potential. Such Coulomb-gas models are ubiquitous in random matrix theory (RMT), where the charges are identified with the eigenvalues of a unitary $N\times N$ matrix in the circular unitary ensemble (CUE). Standard RMT techniques can therefore be adapted to study the large-$N$ behavior of the Loschmidt amplitude in terms of the collective density of charges \cite{eynard_random_2018}.

There is, however, an important difference with respect to standard Coulomb-gas problems. For purely imaginary time $\tau$, the action satisfies the  symmetry condition
\begin{eqnarray}
    \overline{S({\bf q})}=S\big(\overline{\bf q}\big) \ ,
    \label{eq:symmetrie 1}
\end{eqnarray}
which ensures that the action remains real on the real axis.
However, in real time, this symmetry is broken as the external potential is purely imaginary for real momenta $\bm{q}$.
This implies that the large-$N$ saddle point configuration of charges $z_n$ is not constrained to lie on the initial contour  $|z|=1$. 

Assuming that the charges are distributed along an a priori unknown contour $\gamma$  in the complex plane, we introduce the corresponding normalized density 
\begin{equation}
\rho(z)=\lim_{N\to\infty}\frac{1}{N}\sum_{j=1}^N
\delta_\gamma(z,z_j) \quad \text{and}\quad\int_{\gamma}\rho(z)\,|dz|=1,
\label{eq:normalisation densite}
\end{equation}
where $\delta_\gamma$ denotes the Dirac distribution supported on the
contour $\gamma$ that is defined by the property $\int_{\gamma} f(z)\delta(z,a) |\dd z| =  f(a)$ for $a\in \gamma$.
Here $|\dd z|$ is the arc-length element along the contour $\gamma$. For a particular parametrization $s\to z(s)$ it is given by $|\dd z| = |z'(s)|\dd s$.

Following \cite{alvarez_16} we formally rewrite\footnote{This procedure introduces a  global prefactor $C_N = (-1)^N \ii^{-N^2} 2^{-N(N-1)}$ in the Loschmidt amplitude \eqref{eq:LA_rearr} which does not affect the results.}  the Loschmidt amplitude as a functional integral\footnote{Strictly speaking, the Jacobian from the microscopic momenta $\{z_n\}$ to the collective density field $\rho(z)$  produces a subleading contribution to the action of order $O(N)$  and will therefore be neglected.} over charge densities in the plane $z=e^{\ii q}$: 

\begin{equation}
     A_N(t) \simeq 
    \int D[\rho] \, e^{-N^2 S[\rho]} \ ,
    \label{path integral}
\end{equation}
where the action  $S[\rho]$ describes a continuous Coulomb gas
\begin{equation}
    S[\rho] =
    \int_\gamma V(z) \rho(z)\,|dz|
    - \frac{1}{2} \int_{\gamma}\!\int_{\gamma} \log(z - z')^2\rho(z)\rho(z')\,|dz|\,|dz'| \,, 
    \label{eq:action}
\end{equation}
and the external potential is explicitly complex, 
\begin{equation}
    V(z) =  \log z + \ii \tau\left(z + \frac{1}{z}\right) \ .
\end{equation}

\subsection{Saddle point analysis}

The path integral~\eqref{path integral} is dominated in the large-$N$ limit by
its saddle point given by 
\begin{equation}
\frac{\delta}{\delta\rho(z)}
\left[
S[\rho]
+
\mu\left(
1-\int_\gamma \rho(z)\,|dz|
\right)
\right]
=0.
\label{eq:variation}
\end{equation}
It can be written as the electrostatic equilibrium condition
(see Appendix~\ref{App:Electrostatic} for a detailed electrostatic interpretation)
\begin{equation}
    \Re\Phi(z)=\mu,
    \qquad z\in\gamma \, ,
    \label{Anti stockes line}
\end{equation}
which determines the support $\gamma$ as an equipotential contour of the
electrostatic potential $\Phi(z)$ defined by:
\begin{equation}
    \label{eq:holomorphic_potential}
    \Phi(z) =  V(z)
    - \int_\gamma \log(z - z')^2\, \rho(z')\,|dz'| \ .
\end{equation}
In a multi-cut phase where the contour $\gamma = \bigcup_i \gamma_i$, Eq.~\eqref{Anti stockes line} further requires the
electrostatic potential difference between any two occupied cuts to vanish,
\begin{equation}
    \operatorname{Re}\Big(\Phi(z_i)-\Phi(z_j)\Big)=0,
    \qquad z_i\in\gamma_i,\quad z_j\in\gamma_j 
    \label{equ chimique}
\end{equation}
The corresponding filling fractions are then determined by the saddle itself.

Differentiating $\Phi$, we obtain the corresponding electric field $y(z)$,
\begin{equation} 
    \label{eq:spectral_curve}
    y(z)  = V'(z) - 2\omega(z) \ , 
\end{equation}
where $\omega(z)$ is the resolvent, which is analytic on $\mathbb{C} \setminus \{\gamma\}$, 
\begin{equation}
    \label{eq:omega}
   \omega(z)= \mathrm{P.V.} \int_\gamma \frac{\rho(z')\,|dz'|}{z - z'}  \ .
\end{equation}
The density supported on $\gamma$ is then obtained from the discontinuity of the electric field through the Sokhotski--Plemelj formula, with $y_+(z)$ and $y_-(z)$ denoting its boundary values on the two sides of the contour:
\begin{equation}
    \frac{y_{\pm}(z)}{2\pi \ii}\,dz
    =
    \rho(z)\,|dz|,
    \qquad z\in\gamma.
    \label{lien densite et y}
\end{equation}

\subsection{Solution of the electrostatic problem }

In the large-$N$ limit, the resolvent $\omega(z)$ satisfies the so-called loop equation (the derivation is recalled in  Appendix~\ref{App:loop}), which reduces to the following quadratic equation, 
\begin{equation}
    V'(z)\omega(z)-\omega^2(z)=\int_{\gamma}\frac{V'(z) - V'(z')}{z - z'}\rho(z')\,|dz'|\, .
    \label{equ: Loop equation}
\end{equation}
The left-hand side can be written in terms of the electric field (\ref{eq:spectral_curve}), and the right-hand side is a polynomial in $z^{-1}$ depending on two parameters:
\begin{equation}
    y^2=V'^2+\frac{d}{z}+\frac{c}{z^2} \ .
    \label{equ: Loop equation2}
\end{equation}
When $z\to\infty$, (\ref{eq:spectral_curve}) and (\ref{eq:omega}) imply $y(z)\simeq\ii\tau-1/z$, and (\ref{equ: Loop equation2}) fixes explicitly the parameter $d=-4\ii \tau$ using the normalization of the density. This leaves \(c\) as the only parameter to be determined:
\begin{align}
    c(\tau)&= -4\ii \tau\int_\gamma \frac{\rho(z)}{z}\,|dz|
    \label{eq:c_def}\, .
\end{align}
Hence we obtain the following exact expression for the electric field:
\begin{equation}
\label{equ:elec field}
    y(z,c) = \pm\frac{\ii \tau}{ z^2} \sqrt{P(z, c)}\, ,
\end{equation}
where \( P(z, c) \) is the following quartic polynomial 
\begin{equation}
    \label{eq:y2_poly}
    P(z,c) =\qty(z^2+\qty(\frac{\ii-\sqrt{4\tau^2+c}}{\tau})z+1)\qty(z^2+\qty(\frac{\ii+\sqrt{4\tau^2+c}}{\tau})z+1)\, .
\end{equation}
Note that  the symmetry, 
  $  S_{\tau}[-{\bf q}]=S_{\tau}[{\bf q}] \ ,$
implies that the polynomial $P(z,c)$ is palindromic, i.e., it satisfies $P(z)=z^4 P(1/z)$, so that its four roots come in inverse pairs $\{a,1/a,b,1/b\}$.
The structure of this polynomial, and in particular the multiplicity of its roots, controls the topology of the density support $\gamma$, determining whether the support is ungapped, gapped, or split into two components \cite{alvarez_16,eynard_random_2018}.
It is therefore natural to interpret the parameter $c$ as an order parameter of the system. Determining it 
explicitly will be the purpose of the next section.

\subsection{Calculation of the order  parameter}
\label{subsection: order parameter}

The contour  $\gamma$  is not given a priori and has to be determined in
the solution of the problem. Its topological structure will play a key role in identifying the different phases  and it will be  crucial to first analyze the asymptotic regimes in order to anticipate the changes of  $\gamma$'s topology.

\begin{itemize}
\item In the \textbf{extreme long-time regime} $(\tau \to \infty)$, the external potential dominates over the logarithmic interactions. The saddle points then localize along the steepest-descent paths of the external potential \cite{KLM18} around $z=\pm 1$. As a consequence, one expects the density support $\gamma$ to consist of two disjoint symmetric cuts centered around $z=\pm1$.
\item In the \textbf{extreme short-time regime} $(\tau \to 0)$, by contrast, the logarithmic interactions dominate and effectively screen the external potential. The density support therefore consists of a single ungapped cut representing a slightly deformed unit circle.
\end{itemize}

This asymptotic analysis suggests that the dynamical regimes can be distinguished by the topology of the density support $\gamma$ as follows: 
In the short-time regime $(\tau < \tau_c)$, the density support consists of a single ungapped cut, which deforms continuously as $\tau$ increases up to a critical value $\tau_c$. At this point, a transition occurs to the long-time regime $(\tau>\tau_c)$, where the contour $\gamma$ splits into two symmetric disconnected components. As $\tau$ is increased further, these two arcs move continuously toward the steepest-descent paths of the external potential. 
It remains to determine, in each regime, the appropriate order parameter that matches these asymptotic behaviors.

Taking the thermodynamic limit of (\ref{eq:S'}) and taking into account the symmetry $z\leftrightarrow z^{-1}$ of $\rho(z)$, one has the identity
\begin{equation}
    S'(\tau)=2\ii\int_{\gamma}\frac{\rho(z)}{z}\,|dz|\,
\end{equation}
for the dynamical free energy $S(\tau)$, which is then related to the order parameter $c(\tau)$ as
\begin{equation}
  \label{lien Setc}
    S'(\tau) = -\frac{c(\tau)}{2\tau}\,.
\end{equation}
In the short-time regime ($\tau<\tau_c$), the quartic polynomial \eqref{eq:y2_poly} must degenerate into a perfect square in order to produce a single ungapped support \cite{alvarez_16,eynard_random_2018}. This condition gives
\begin{equation}
    c(\tau) = -4\tau^2 \qquad \text{for } \tau\leq\tau_c \, .
    \label{critique dipole}
\end{equation}
In the long-time regime ($\tau>\tau_c$), however, the above expression is no longer valid. This reflects the change in the topology of the support, and another criterion is needed to determine \(c(\tau)\). Indeed, \eqref{critique dipole} is unbounded as $\tau \to \infty$, whereas the correct asymptotic behavior is known to be~\cite{KLM18}
\begin{equation}
\lim_{\tau \to \infty} c(\tau) = -\frac{1}{2}.
\label{dip infini}
\end{equation}
In this regime, the endpoints of the support are fully determined by the four branch points $\{a,1/a,b,1/b\}$ of the electric field $y(z,c)$.
Using \eqref{lien densite et y},
the filling fraction carried by the first cut (or population of charges) is
\begin{equation}
    P_1
    =
    \int_{1/a}^{a}
    \frac{dz}{2\pi\ii}\,y(z,c).
\end{equation}
Since both the electric field and the support are complex, the filling
fractions are not guaranteed a priori to be real. Therefore a physically admissible
saddle must give real populations on the different
components of the support. Since the total normalization imposes
$P_1+P_2=1$, it is sufficient to require that one filling fraction must be
real:
\begin{equation}
    \Im\!\left(
        \int_{1/a}^{a}
        \frac{dz}{2\pi\ii}\,y(z,c)
    \right)
    =0.
    \label{eq:dipole}
\end{equation}
However, since the order parameter $c$ is generally complex, an additional condition is required to determine it completely. This is provided by the electrostatic equilibrium condition \eqref{equ chimique} between the two cuts.
These two conditions \eqref{eq:dipole} and \eqref{equ chimique} may admit several branches of solutions for $c$,
corresponding to distinct admissible saddle-point configurations. Identifying
the dominant branch therefore requires additional global information, which can be
obtained by analyzing the asymptotic regimes and the behavior of the solution
across the transition.

In the present real-time problem, the large-$\tau$ asymptotics already provide
this information: the dominant configuration is known to be symmetric
\cite{KLM18}. We therefore restrict ourselves to the symmetric branch satisfying
\[
    a=-\bar b.
\]
This condition implies that $c(\tau)$ is real and
$c(\tau)>-4\tau^2$. Moreover, since the two cuts are exchanged by this
symmetry, the condition
\eqref{equ chimique} is automatically satisfied. Hence, only
Eq.~\eqref{eq:dipole} is needed to determine
$c(\tau)$.

\begin{figure}[H]
\centering
\includegraphics[width=1\textwidth]{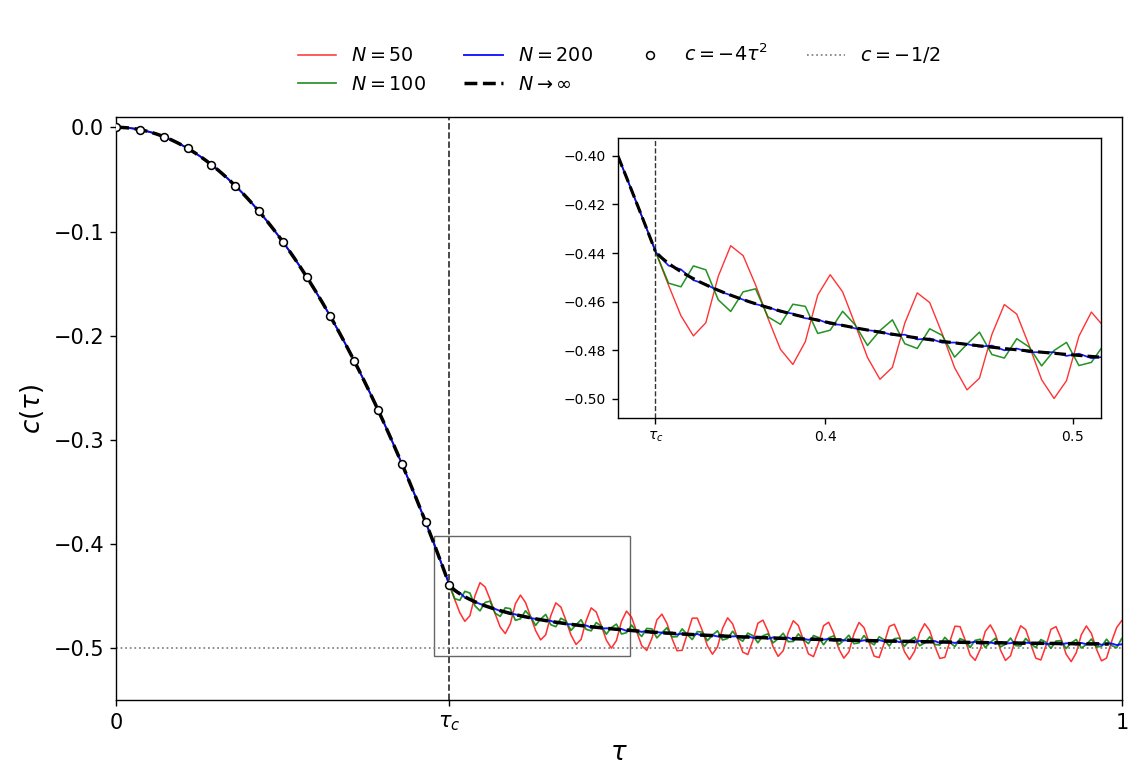}
\caption{Order parameter $c(\tau)$. The colored curves are extracted from the
finite-$N$ Loschmidt amplitude computed through the Toeplitz determinant
representation~\eqref{eq:toeplitz}. The
black dashed curve is the large-N solution obtained from
Eq.~\eqref{eq:dipole}. The open circles indicate the exact short-time
result $c(\tau)=-4\tau^2$, while the grey dotted line denotes the
large-time asymptote $c(\infty)=-1/2$. The inset highlights the convergence
of the finite-$N$ results near the critical point.}
\label{figure dipole}
\end{figure}

Using the change of variable $u=z+z^{-1}$ and introducing the parameter
\begin{equation}
\theta(\tau)
=
\tau^{-1}\sqrt{4\tau^2+c(\tau)}\, ,
\label{theta:def}
\end{equation}
Eq.~\eqref{eq:dipole} can be rewritten as
\begin{equation}
\int_{1/a}^{a}\frac{dz}{2\pi\ii}\,y(z)
=
\frac{\tau}{\pi}
\int_{2}^{\theta-\frac{\ii}{\tau}}du\,
\sqrt{
\frac{
\left(u+\frac{\ii}{\tau}+\theta\right)
\left(u+\frac{\ii}{\tau}-\theta\right)
}{
(u-2)(u+2)
}
}.
\label{E:elliptique}
\end{equation}
With this parametrization, the
integral on the right-hand side of \eqref{E:elliptique} can be reduced to a linear combination of the
complete elliptic integrals $K(k)$, $E(k)$, and $\Pi(n,k)$
\cite{Abramowitz1964,Byrd} (see also Appendix~\ref{App:Elliptic}),
\begin{align}
 \int_2^{\theta-\frac{\ii}{\tau}} du
    \sqrt{ \frac{(u + \frac{\ii}{\tau} + \theta)(u + \frac{\ii}{\tau} - \theta)}{(u - 2)(u + 2)} }
    &= A(\theta,\tau)\, E(k)
    +B(\theta,\tau)\, K(k)
   +C(\theta,\tau)\, \Pi(n,k),
   \label{Decomp_elliptique}
\end{align}
with coefficients 
\begin{equation}
\begin{aligned}
A(\theta,\tau)
&= \ii\sqrt{
\left(\theta+2+\frac{\ii}{\tau}\right)
\left(\theta+2-\frac{\ii}{\tau}\right)
}, \\[4pt]
B(\theta,\tau)
&= -2\left(\theta+\frac{\ii}{\tau}\right)
\sqrt{
\frac{\frac{\ii}{\tau}-\theta-2}
     {\frac{\ii}{\tau}+\theta+2}
}, \\[4pt]
C(\theta,\tau)
&= -\frac{8}{
\tau\sqrt{
\left(\theta+2+\frac{\ii}{\tau}\right)
\left(\theta+2-\frac{\ii}{\tau}\right)
}
}\, .
\end{aligned}
\end{equation}
The elliptic modulus \(k\) and the characteristic \(n\) of the elliptic integral of the third kind are 
\begin{equation}
    k=\sqrt\frac{\left(\theta-\frac{\ii}{\tau}-2\right)\left(\theta+\frac{\ii}{\tau}-2\right)}{\left(\theta-\frac{\ii}{\tau}+2\right)\left(\theta+\frac{\ii}{\tau}+2\right)} \ ,
    \qquad
    n =\frac{\theta-\frac{\ii}{\tau}-2}{\theta-\frac{\ii}{\tau}+2} \ .
    \label{def: k alpha}
\end{equation}
The resulting transcendental equation \eqref{eq:dipole} does not admit a closed-form solution for $c(\tau)$ in the long-time regime. Nevertheless, it can be solved numerically (see Fig.~\ref{figure dipole}), and it allows one to extract the exact scaling of the dynamical free energy $S(\tau)$ near the critical point (see Sec.~\ref{section:Critical Regime}).
As an independent check, we extract the order parameter directly from
the finite-$N$ Loschmidt amplitude computed through the Toeplitz
determinant representation~\eqref{eq:toeplitz} and using
Eq.~\eqref{lien Setc}. As shown in Fig.~\ref{figure dipole}, the resulting
curves converge toward the large-N solution obtained independently
from Eq.~\eqref{eq:dipole}. This solution also reproduces as it should the exact
short-time result $c(\tau)=-4\tau^2$ and approaches the expected
large-time limit $c(\infty)=-1/2$.

\section{DQPT: topological transition between the unbroken-heart and broken-heart phases}
\label{sec:DQPT}

\subsection{Short-time regime (unbroken-heart phase)}
\label{sect:unbroken heart phase}

\begin{figure}[h]
    \centering
    \includegraphics[width=1.\textwidth]{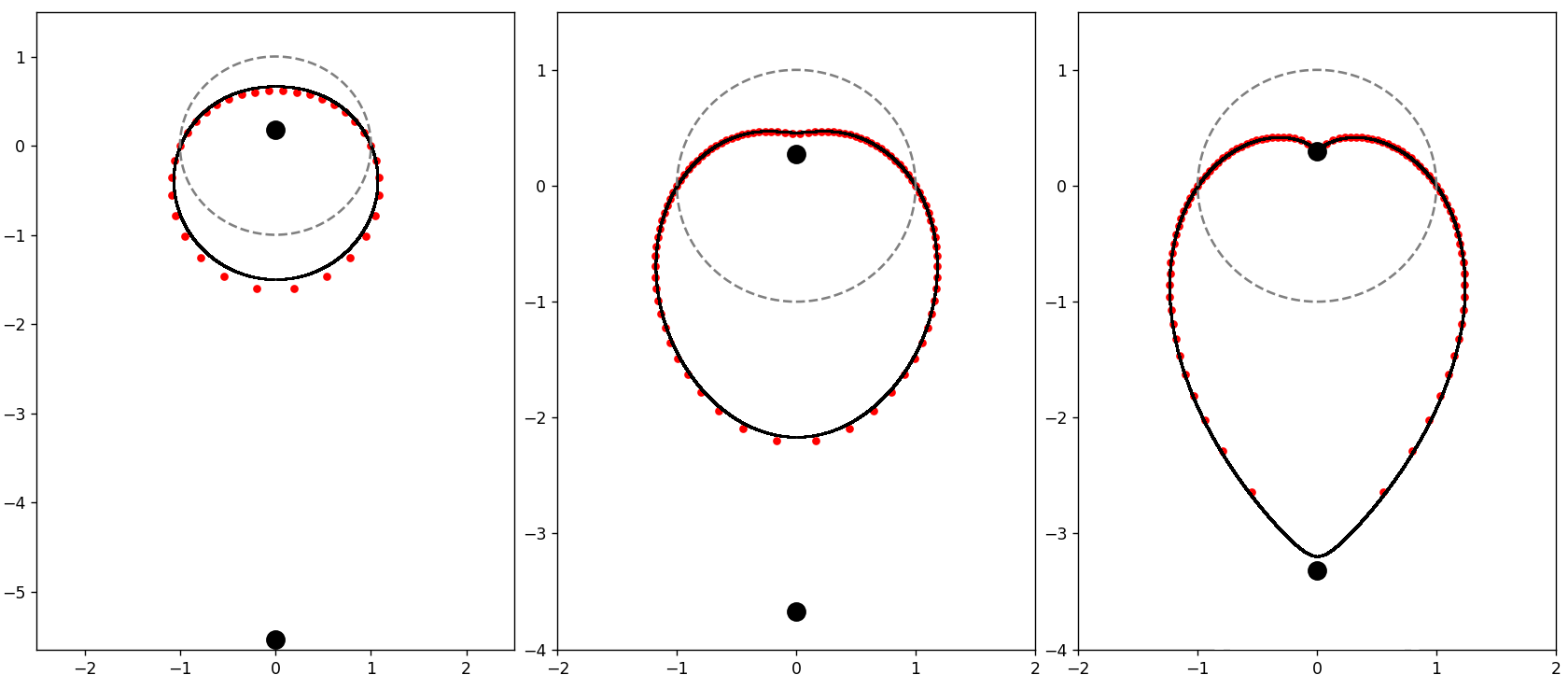}
\caption{
Evolution of the dominant instanton configuration in the short-time
regime in the complex plane $z=e^{iq}$. From left to right:
$\tau=0.25$, $\tau=0.29$, and $\tau\simeq\tau_c=0.331$, with
$N=40$, $N=90$, and $N=90$, respectively. Red dots show the discrete
saddle points obtained by solving numerically \eqref{eq:saddle point q}, while the black curve
represents the continuum support $\gamma$ determined exactly in
\eqref{parametrage gamma}. Black dots indicate the zeros of $y(z)$, and the grey dotted
curve is the unit circle. As $\tau$ approaches $\tau_c$, the zeros move
toward the support, which continuously deforms into the unbroken-heart
configuration at the transition.
}
\label{fig:configurations saddles reélle }
\end{figure}

Substituting Eq.~\eqref{critique dipole} into~\eqref{eq:y2_poly}, the
electric field simplifies to
\begin{equation}
y(z)=\pm \frac{\ii \tau}{ z^2}\left(z^2+\frac{\ii}{\tau}z+1\right) \ .
\end{equation}
Integrating this expression, we obtain the electrostatic potential 
\begin{equation}
\Phi(z)=\pm\Big(\ii \tau\left(z-z^{-1}\right)-\log z\Big) \ .
\label{def :Wh}
\end{equation}
The density support $\gamma$ is then determined by solving\footnote{The saddle points of the external potential, $z=\pm 1$, are fixed points of the support and can  be used to determine the chemical potential  $\mu=\operatorname{Re}\Phi(\pm 1)=0$.} the equilibrium condition~\eqref{Anti stockes line}, which leads to the following parametric representation of the contour:
\begin{equation}
    z_{\gamma}(r) = r \exp\!\Big( -i \arcsin\big( \frac{ r \log r}{\tau(r^2 + 1)} \big) \Big)
    \qquad \text{for } ~\left| \frac{r \log r}{\tau(r^2 + 1)} \right| < 1 \ .
    \label{parametrage gamma}
\end{equation}
Figure~\ref{fig:configurations saddles reélle } illustrates how the contour $\gamma$ evolves dynamically as $\tau$ evolves in the short-time regime.
In the limit $\tau \to \tau_c$, the contour deforms continuously until it takes the shape of a heart, which we referred to as the unbroken-heart.

The large black dots correspond to the points $(z_0\,,\,1/z_0)$ where the electric field $y(z)$ vanishes:
\begin{equation}
z_0(\tau)=\frac{\ii\left(-1 + \sqrt{1+4\tau^2}\right)}{2\tau} \, .
\end{equation}
These zeros lie outside the support in this regime. As $\tau$ increases, they move towards the contour from both sides and eventually touch it at two distinct points along the imaginary axis. At the critical time $\tau_c$ where this occurs, the density vanishes at the two touching points, as follows from \eqref{lien densite et y}. This signals the breakdown of the ungapped one-cut solution: beyond this point, the support must split into two disconnected components. Hence, the condition that a zero of $y(z)$ lies on the support provides a criterion for determining $\tau_c$:
\begin{align}
\operatorname{Re}\Phi\big(z_0(\tau_c)\big)=0 \Leftrightarrow 
\sqrt{1+4\tau^2_c}+\log\left(\frac{-1 + \sqrt{1+4\tau_c^2}}{2\tau_c}\right)=0 \ .
\label{eq: lbd critique}
\end{align}
This gives $\tau_c\approx0.331$ in agreement with Ref.~\cite{perez-garcia}. See also Appendix~\ref{Appendix Recursion} for an alternative derivation
of this value using  the  exact recursion relation satisfied by the Toeplitz determinant~\niceref{eq:toeplitz}.
Finally, using \eqref{critique dipole} and \eqref{lien Setc}, we obtain the dynamical free energy in the short-time regime
\begin{equation}
    S(\tau) = \tau^2 \qquad (\tau \leq \tau_c)\ \ ,
\end{equation}
leading to the Gaussian decay \niceref{eq:LE_short}

\subsection{Long-time regime (broken-heart phase)}
\label{Sec : Broken Heart Phase}
\begin{figure}[h]
    \centering
    \includegraphics[width=1\textwidth]{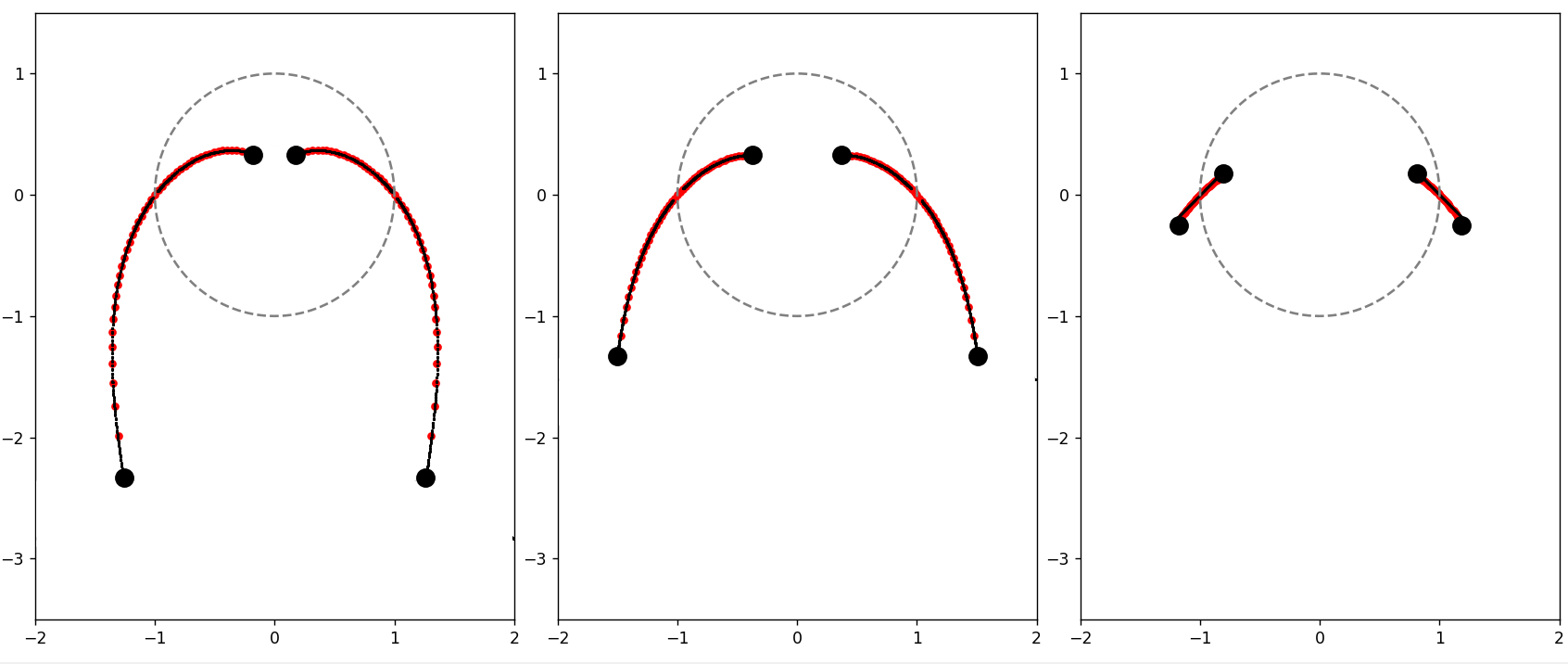}
\caption{
Evolution of the dominant instanton configuration in the long-time
regime. From left to right: $\tau=0.5$, $\tau=1$, and $\tau=10$, with
$N=90$. Red dots show the discrete saddle points obtained by solving numerically
\eqref{eq:saddle point q}, while the black curves represent the continuum support
$\gamma$ determined from \eqref{Anti stockes line}. Black dots indicate the branch
points of $y(z)$ and the endpoints of the support. The grey dotted curve is the unit circle. In this
regime, the support consists of two symmetric disconnected cuts forming
the broken-heart configuration. As $\tau$ increases, the two cuts
continuously approach the steepest-descent contours of the external
potential.
}
\label{fig:configurations long time}
\end{figure}

In this regime, the order parameter $c(\tau)$ does not admit a closed-form expression and is determined numerically from Eq.~\eqref{eq:dipole} (see Fig.~\ref{figure dipole}). The corresponding electrostatic potential $\Phi(z)$ can also be expressed in terms of incomplete elliptic integrals, and the equilibrium condition~\eqref{Anti stockes line} is then solved numerically to determine the contour $\gamma$.

Figure~\ref{fig:configurations long time} illustrates how the contour $\gamma$
evolves dynamically with $\tau$ in the long-time regime. In this regime, the unbroken-heart of the short-time regime breaks into two disconnected and symmetric cuts, which we refer to as the broken- heart. As $\tau$ further increases, the two arcs continuously approach the steepest-descent paths of the external potential, as expected from Ref.~\cite{KLM18}.

Finally, as a further consistency check, substituting the numerical solution $c(\tau)$ of
Eq.~\eqref{eq:dipole} into the filling fraction, we do find $P(\tau\geq\tau_c)=
    \frac{1}{2}$
within numerical precision.
This confirms that, in real time, the dominant solution remains symmetric
throughout the long-time regime and that no charge transfer occurs between
the two cuts as $\tau$ varies. This is also independently supported by
Fig.~\ref{figure dipole}, where the order parameter obtained by imposing
the symmetry condition together with Eq.~\eqref{eq:dipole} agrees with the
finite-$N$ results extracted from the Toeplitz determinant.
The equal population of the two cuts can also be verified analytically at the
transition point. Since the explicit short-time solution remains valid at
$\tau=\tau_c$, using Eq.~\eqref{def :Wh} together with
Eq.~\eqref{eq: lbd critique} gives
\begin{equation}
P(\tau_c) \;=\; \frac{1}{\pi}\,\operatorname{Im}
\Phi\big(z_0(\tau_c)\big)
\;=\;\frac{1}{2} \ .
\label{eq:Pc_half}
\end{equation}

\subsection{Critical regime: broken-heart transition and birth of a cut}
\label{section:Critical Regime}

The phase transition is characterized by a change in the topology of
the density support, from a single ungapped cut to two symmetric
disconnected cuts. In random matrix theory, such a one-cut to two-cut
transition is known as a birth-of-a-cut transition
\cite{eynardUniversalDistributionRandom2006,
moRiemannHilbertApproach2008,
claeysBirthCutUnitary2008} and is typically accompanied by logarithmic
singularities in the free energy.

To determine the critical scaling, we start from the filling-fraction
condition~\eqref{eq:dipole}. As $\tau\to\tau_c^+$, the parameter
$\theta(\tau)$ defined in Eq.~\eqref{theta:def} tends to zero, while the
elliptic modulus $k$ approaches one. 
In this limit, the asymptotic expansions of the elliptic integrals contain logarithmic terms (see Appendix~\ref{App:Elliptic}).
Hence using these asymptotic expansions, we expand the elliptic integrals in
Eq.~\eqref{Decomp_elliptique}, together with the coefficients
$A(\theta,\tau)$, $B(\theta,\tau)$, and $C(\theta,\tau)$, up to second
order in $\theta$.
Although the intermediate expressions are lengthy, they simplify
considerably once all contributions are combined. In particular, the
isolated term proportional to $\log\theta$ cancels exactly, while the
term linear in $\theta$ vanishes by symmetry.
We then expand the remaining coefficients to first order in
$\tau-\tau_c$, approaching the transition from the long-time regime,
$\tau\to\tau_c^+$. After carrying out these expansions, the constant
term at zeroth order vanishes exactly at $\tau=\tau_c$, yielding a
condition equivalent to the criticality condition \eqref{eq: lbd critique}. Therefore the leading
non-vanishing terms of the condition \eqref{eq:dipole} give
\begin{equation}
    \theta^2\log\theta\simeq
    \frac{-2(1+4\tau_c^2)}{\tau_c^3}(\tau-\tau_c).
    \label{eq:theta_critical_scaling}
\end{equation}
Using the leading-logarithmic approximation
$\log\theta\sim\frac{1}{2}\log(\tau-\tau_c)$ in this transcendental
relation, and then using Eq.~\eqref{theta:def}, we obtain
\begin{equation}
    c(\tau)=-4\tau^2
-\frac{4(1+4\tau_c^2)}{\tau_c}\frac{(\tau-\tau_c)}{\log(\tau-\tau_c)},
    \qquad
    \tau\to\tau_c^+.
    \label{eq:critical_c_scaling}
\end{equation}
Substituting  Eq.~\eqref{eq:critical_c_scaling} into \eqref{lien Setc}, one obtains the result \eqref{eq:derive action}.

Consequently, $S(\tau)$ and its first two derivatives remain continuous
at $\tau=\tau_c$, whereas the third derivative diverges as
$\tau\to\tau_c^+$. This characterizes a phase transition of order
$2+0$: it is logarithmically weaker than a standard second-order
transition, but more singular than the third-order GWW transition
arising in imaginary time, for which the third derivative displays a
finite discontinuity.

\subsection{Relation to the GWW model}

Under a Wick rotation to imaginary time, the Loschmidt amplitude defined in
Eq.~\eqref{eq:LA_rearr} coincides exactly with the partition function of the
Gross-Witten-Wadia (GWW) model~\cite{GrossWitten1980,WADIA}, a well-known
model in lattice gauge theory exhibiting a third-order phase transition in the
large-$N$ scaling limit at $\tau_c=1/2$. However, the real-time asymptotics of
the Loschmidt echo cannot be inferred from a naive Wick rotation of the GWW
solution. Although the short-time dynamical free energy is related to its
imaginary-time counterpart by analytic continuation, the critical points and
the long-time phases are different. More importantly, the nature of the
transition itself is different.

In imaginary time, the action defined in Eq.~\eqref{eq:LA_act} is real-valued
and obeys the conjugation symmetry~\eqref{eq:symmetrie 1}. As a consequence,
the density support $\gamma$ remains on the unit circle, and the problem reduces
to determining a positive density $\rho(z)$ along this fixed contour. For
$\tau\leq\tau_c$, the density remains positive on the full unit circle. For
$\tau>\tau_c$, no positive solution exists on the full circle, and the support
must instead be restricted to an open arc of the unit circle.
\begin{figure}[H]
    \centering
    \includegraphics[width=1.\textwidth]{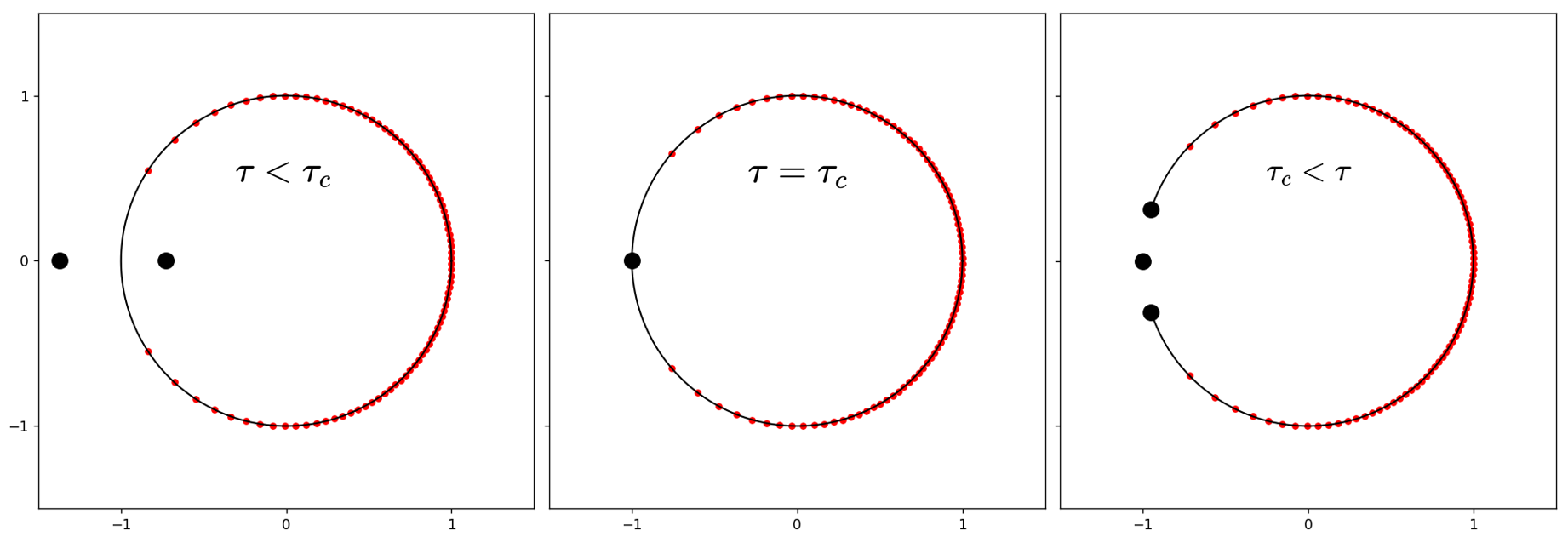}

\caption{
Evolution of the dominant saddle-point configuration in imaginary time.
From left to right: $\tau=0.47$, $\tau=\tau_c=0.5$, and $\tau=0.51$,
with $N=90$. Red dots show the discrete saddle points obtained by
solving numerically \eqref{eq:saddle point q}, while the black curves represent the continuum
support $\gamma$ determined from \eqref{Anti stockes line}. Black dots indicate
the zeros of $y(z)$. Below the transition, the support is the full unit circle.
At $\tau=\tau_c$, the zero at $z=-1$ becomes double, and for
$\tau>\tau_c$ a gap opens around this point, leaving a single gapped arc.
}
\label{fig:GWW}
\end{figure}
Thus, in imaginary time, the transition is produced by the opening of a gap
within a single connected support, rather than by the birth of an additional
cut, as illustrated in Fig.~\ref{fig:GWW}. This difference of topology
explains why the logarithmic singularity of the real-time transition is absent.

An alternative way to understand this distinction is to consider the
condition~\eqref{eq:dipole}. In imaginary time, the one gapped cut configuration in the long-time regime is obtained by requiring one inverse pair of roots of the polynomial
\eqref{eq:y2_poly}  to coalesce into a double root~\cite{alvarez_16}. Since the
polynomial is then real, the two remaining roots form a complex-conjugate pair
and determine the endpoints of the support. The one-cut configuration is
therefore completely fixed by its endpoints, and the condition
\eqref{eq:dipole} is automatically satisfied.

\section{Loschmidt echo in complex time and post-selection}
\label{sec:complex_time}

As discussed in Sec.~\ref{section complex times post selection}, post-selection of the work
distribution naturally leads to the Loschmidt amplitude evaluated at complex
time. We therefore extend the saddle-point analysis of the previous sections to
complex time $t=|t|e^{i\phi}$, with $0\leq\phi\leq\frac{\pi}{2}$, and consider
the generalized Loschmidt amplitude
\begin{equation}
    A_N(t,\phi)
    =
    \langle \Psi_0 |
    e^{-\ii |t|e^{\ii\phi}H}
    |\Psi_0\rangle \, .
    \label{eq:complex_loschmidt_amplitude}
\end{equation}
We will show that the complex-time evolution is governed by three distinct
phases, characterized by different topologies of the dominant support: an
ungapped one-cut phase, a two-cut phase, and a gapped one-cut phase. 

\subsection{Short-time regime}
\label{sec:complex_short_time}

\begin{figure}[H]
    \centering
\includegraphics[width=1.\textwidth]{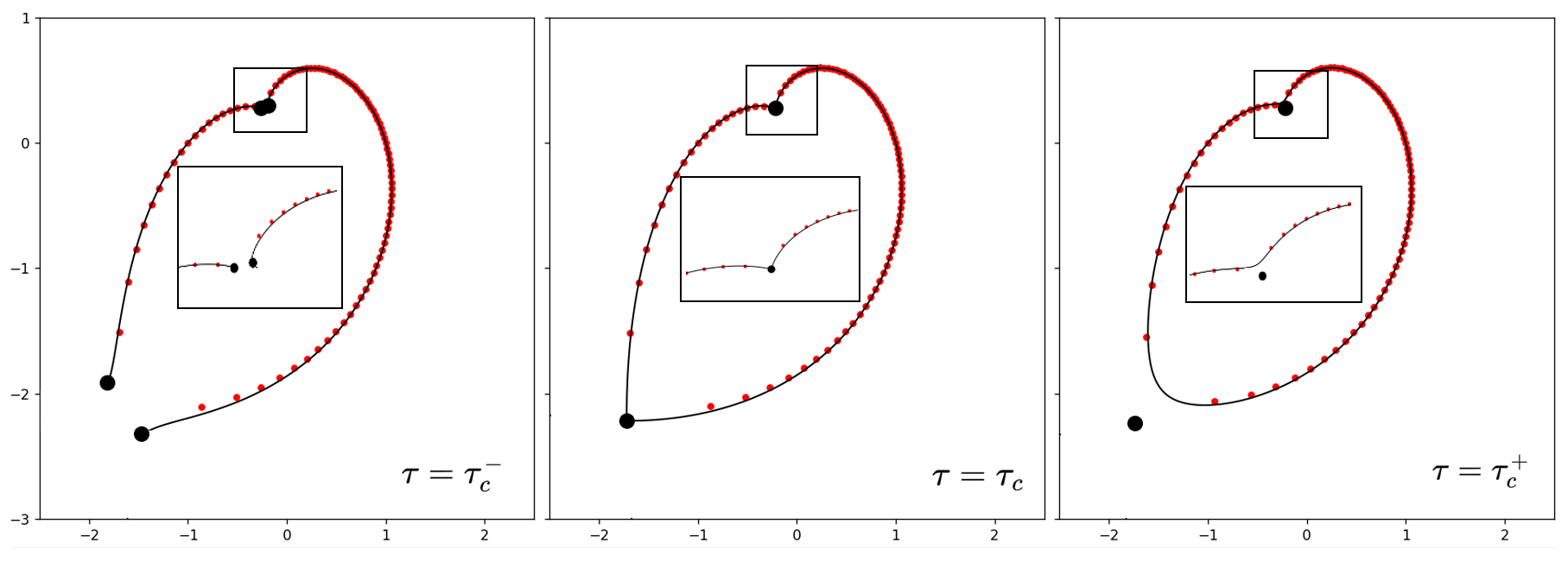}
\caption{
Topological transition of the dominant instanton configuration across the first critical line at
$\phi=\pi/4$ in the complex $z$-plane. From left to right:
$\tau=0.36$, $\tau=0.364$, and $\tau=0.37$, with $N=90$. Red dots show the discrete saddle
points obtained by solving numerically \eqref{eq:saddle point q}, while the black curves represent
the continuum support $\gamma$ determined from \eqref{Anti stockes line}. Black dots
indicate the zeros of $y(z)$. 
}
 \label{fig:Complexe }
 \end{figure}
 The short-time analysis follows directly the strategy developed in
Sec.~\ref{sect:unbroken heart phase}, by changing $\tau \longrightarrow \tau e^{i\phi}$.
The electric field therefore reads
\begin{equation}
    y(z,c)
    =
    \pm
    \frac{i\tau e^{i\phi}}{z^2}
    \sqrt{P(z,c)}
    \label{def: y complexe}
\end{equation}
where
\begin{align}
    P(z,c)
    &= \left(z^2
        +
        \left(
            \frac{i-\sqrt{4\tau^2 e^{2i\phi}+c}}
                 {\tau e^{i\phi}}
        \right)z
        +1\right)\left(
        z^2
        +\left(
            \frac{i+\sqrt{4\tau^2 e^{2i\phi}+c}}
                 {\tau e^{i\phi}}\right)z
        +1\right)
        \label{equ: poly complexe}
\end{align}
As in the real-time problem, the ungapped solution is obtained by requiring
the quartic polynomial to degenerate into a perfect square which fixes the
order parameter to
\begin{equation}
    c(\tau,\phi)
    =
    -4\tau^2 e^{2i\phi}.
    \label{dipole complexe short}
\end{equation}
The corresponding electrostatic potential is
\begin{equation}
    \Phi(z)
    =
    \pm
    \left(
        i\tau e^{i\phi}
        \left(z-z^{-1}\right)
        -
        \log z
    \right).
\end{equation}
The density support $\gamma$ is then deduced by solving the electrostatic equilibrium condition~\eqref{Anti stockes line} and is shown in Fig.~\ref{fig:Complexe } for $\phi=\pi/4$.

Following the same argument as in Eq.~\eqref{eq: lbd critique}, we deduce the first critical line $\tau_{c_1}(\phi)$:
\begin{equation}
    \operatorname{Re}
    \left(
        \sqrt{1+4\tau_{c_1}^2e^{2i\phi}}
        +
        \log
        \left(
            \frac{
                -1+\sqrt{1+4\tau_{c_1}^2e^{2i\phi}}
            }{
                2\tau_{c_1}e^{i\phi}
            }
        \right)
    \right)
    =0.
    \label{eq:first_complex_boundary_explicit}
\end{equation}

\subsection{Long-time regime}
\label{sec:complex_long_time}

\subsubsection{Intermediate phase }
Immediately after crossing the first critical line $\tau>\tau_{c_1}(\phi)$, the topology of the new
saddle can be inferred from its filling fractions. Since the solution on the
critical line is known explicitly, the population carried by one of the two
emerging components is
\begin{equation}
  P_c(\phi)\;=\; \frac{1}{\pi}\,\operatorname{Im}\Phi\left(\frac{\ii\left(-1 + \sqrt{1+4\tau_{c_1}^2e^{2i\phi}}\right)}{2\tau_{c_1} e^{i\phi}}\right) .
  \label{equ: pop critique complexe}
\end{equation}
\begin{figure}[H]
    \centering
    \includegraphics[width=0.7\textwidth]{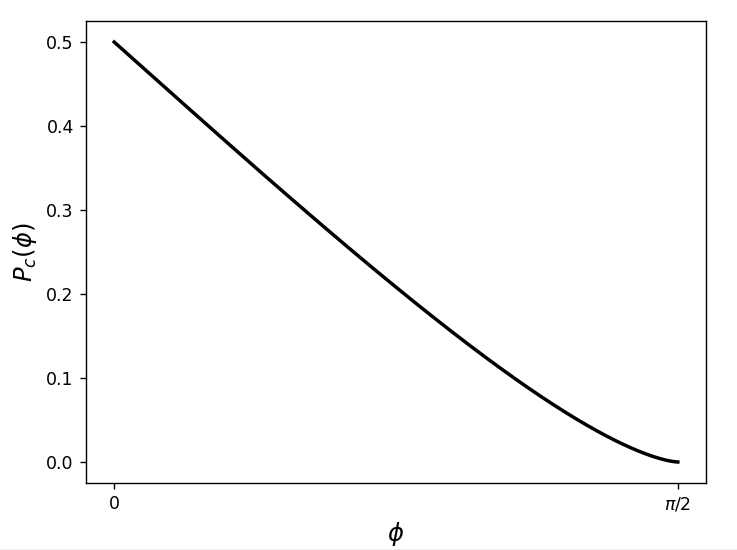}
    \caption{Population \(P_c(\phi)\) carried by one of the two cuts at the first critical line \(\tau_{c_1}(\phi)\).}
    \label{fig: Pop}
\end{figure}

As shown in Fig.~\ref{fig: Pop}, $P_c(\phi)$ decreases continuously from $1/2$ at $\phi=0$ to zero at $\phi=\pi/2$. 
Therefore, for $0\leq\phi<\pi/2$, the solution emerging  across the critical line $\tau_{c_1}(\phi)$ is a two-cut configuration,
with unequal populations carried by its two components. The real-time point
$\phi=0$ is special because the two cuts are related by symmetry and remain
equally populated. The imaginary-time endpoint $\phi=\pi/2$ is also special since
the population of the second component vanishes and the transition reduces to a gapped one-cut solution.

This interpolation already suggests that in complex time, the filling
fractions do not have to remain constant throughout the long-time regime. In
particular, to go from the two-cut configuration for some $\phi<\pi/2$ to the one-cut
solution at $\phi=\pi/2$ requires a redistribution of charges between the two
components.

However, $P_c(\phi)$ only determines the populations immediately
after crossing the first critical line and therefore does not tell us how
they evolve as $\tau$ is increased further at fixed $\phi$. 
It is thus natural to ask whether charge transfer takes place deeper in the long-time
regime, and in particular whether one of the two cuts can eventually become
completely depleted.

A hint already comes from the extreme long-time regime $\tau\to\infty$. Indeed, the contribution of the external potential
to the real part of the action is
\begin{equation}
    \operatorname{Re}
    \left(
        2i\tau e^{i\phi}\cos q
    \right)
    =
    -2\tau\sin\phi\,\cos q.
    \label{eq:complex_external_real}
\end{equation}
For any $\phi>0$, the two stationary points $q=0$ and $q=\pi$ are no longer equivalent. Moreover, this
asymmetry grows linearly with $\tau$. This suggests that, sufficiently far in
the long-time regime, the cut around $q=0$ should become energetically
favored over the other, eventually leading to a one-cut configuration. 
Indeed, we show in the next section that there exists a second critical line $\tau_{c_2}(\phi)$, along which one of the two cuts becomes completely depleted and dies.

\subsubsection{Second transition: death of a cut}
\label{sec:second_boundary}

The second critical line $\tau_{c_2}(\phi)$ is determined by the degeneracy condition at which the two-cut solution collapses onto the one-cut solution.

A one-cut configuration is obtained  when one inverse pair of branch points of the electric field \eqref{def: y complexe}
coalesces \cite{alvarez_16,eynard_random_2018}. This condition is satisfied for
\begin{equation}
    c_{\mathrm{1-cut}}(\tau,\phi)
    =
    -1-4i\tau e^{i\phi}.
    \label{eq:c_one_cut_complex}
\end{equation}
For this value, two branch points coalesce at $z=-1$, and the electric field
reduces to
\begin{equation}
    y_{\mathrm{1-cut}}(z)
    =
    \pm
    \frac{
        i\tau e^{i\phi}(z+1)
    }{
        z^2
    }
    \sqrt{P_{\mathrm{1-cut}}(z)},
    \label{eq:y_one_cut_complex}
\end{equation}
where
\begin{equation}
    P_{\mathrm{1-cut}}(z)
    =
    z^2
    +
    2
    \left(
        -1+\frac{ie^{-i\phi}}{\tau}
    \right)z
    +1.
    \label{eq:P_one_cut_complex}
\end{equation}
The two remaining branch points are
$\left(a(\tau,\phi),1/a(\tau,\phi)\right)$, where
\begin{equation}
    a(\tau,\phi)
    =
    \left(
        1-\frac{\ii e^{-\ii\phi}}{\tau}
    \right)
    +
    \sqrt{
        \left(
            1-\frac{\ii e^{-\ii\phi}}{\tau}
        \right)^2
        -1
    }.
    \label{eq:a_one_cut_complex}
\end{equation}
Integrating Eq.~\eqref{eq:y_one_cut_complex}, we obtain the corresponding electrostatic potential of this configuration: 
\begin{align}
    \Phi_{\mathrm{1-cut}}(z)
    &=
    \pm
    \left(
        i\tau e^{i\phi}
        \left(
            1-\frac{1}{z}
        \right)
        \sqrt{P_{\mathrm{1-cut}}(z)}
        +
        \log
        \left(
            \frac{
                1+
                \left(
                    -1+\frac{ie^{-i\phi}}{\tau}
                \right)z
                +
                \sqrt{P_{\mathrm{1-cut}}(z)}
            }{z^2
-z\left(
    1-\frac{\ii e^{-\ii\phi}}{\tau}
\right)
+z\sqrt{P_{\mathrm{1-cut}}(z)}}
        \right)
    \right).
    \label{eq:phi_one_cut_complex}
\end{align}
Approaching the critical line from the two-cut phase, one of the filling fractions
vanishes and the corresponding cut collapses to the double root $z=-1$.
At the boundary, the resulting one-cut configuration is degenerate and must therefore still satisfy
the equilibrium condition Eq.~\eqref{equ chimique} that characterizes the two-cut phase. 
\begin{figure}[h]
\centering
\includegraphics[width=1.\textwidth]{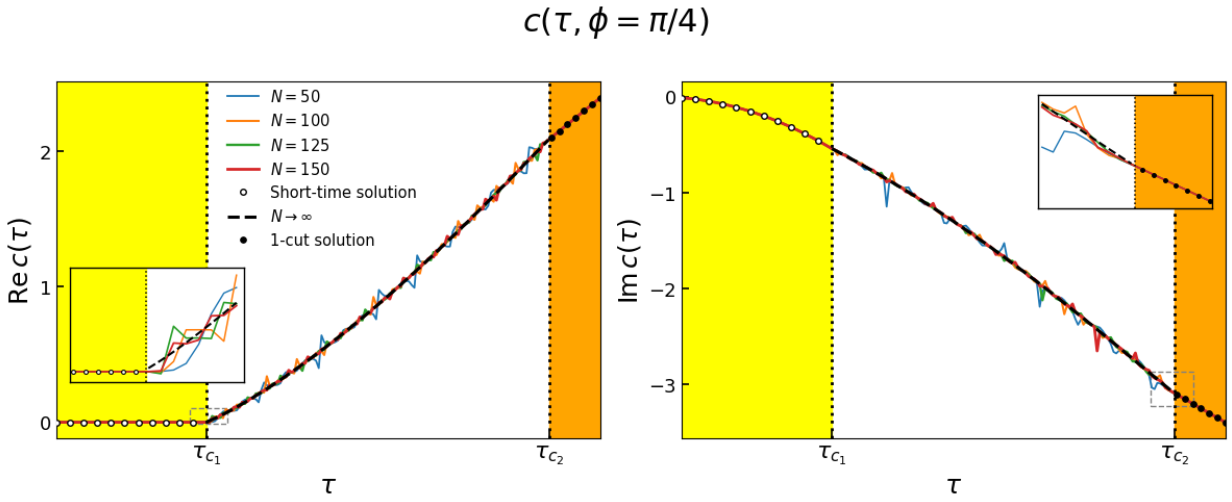}
\caption{
Order parameter $c(\tau,\phi)$ at $\phi=\pi/4$. The left and right panels show
$\operatorname{Re}c(\tau)$ and $\operatorname{Im}c(\tau)$, extracted from the finite-$N$
Toeplitz determinant for $N=50,100,125$, and $150$. Open circles denote the exact
short-time solution \eqref{dipole complexe short}, while filled black circles indicate the
gapped one-cut solution \eqref{eq:c_one_cut_complex}. The dashed black line shows the
$N\to\infty$ solution in the two-cut phase, obtained by solving
Eqs.~\eqref{eq:dipole} and \eqref{equ chimique}. The yellow and orange shaded
regions indicate the ungapped and gapped one-cut phases predicted by the phase diagram.
The insets show enlarged views of the vicinity of the two critical points.
}
\label{figure dipole complex}
\end{figure}
Using the one-cut electrostatic potential
\eqref{eq:phi_one_cut_complex}, this happens for
\begin{equation}
    \operatorname{Re}
    \Phi_{\mathrm{1-cut}}(-1)
    =
    \operatorname{Re}
    \Phi_{\mathrm{1-cut}}
    \left(a\left(\tau_{c_2},\phi\right)\right),
    \label{eq:second_complex_boundary}
\end{equation}
which finally determines the second critical line $\tau_{c_2}(\phi)$: 
\begin{equation}
    \operatorname{Re}
    \left(
        2i\tau_{c_2}e^{i\phi}
        \sqrt{
            1-\frac{i}{2\tau_{c_2}e^{i\phi}}
        }
        +
        \operatorname{arcosh}
            \sqrt{
                1+2i\tau_{c_2}e^{i\phi}
            }
    \right)
    =0.
    \label{eq:second_complex_boundary_explicit}
\end{equation}
This critical boundary separates the intermediate two-cut phase from the long-time gapped one-cut phase (see Fig.~\ref{fig:phase_diagram}).

Figure~\ref{figure dipole complex} shows, for $\phi=\pi/4$, that the
finite-$N$ order parameter extracted from the Toeplitz determinant clearly
reproduces the two critical points,
$\tau_{c_1}\simeq 0.364$ and $\tau_{c_2}\simeq 1.092$,
predicted by Eqs.~\eqref{eq:first_complex_boundary_explicit} and
\eqref{eq:second_complex_boundary_explicit}, respectively.
The finite-$N$ results also converge remarkably well towards our
large-$N$ solution in all three phases, thereby providing an independent validation of the
phase diagram obtained from our saddle-point analysis.

\subsubsection{Order of the second transition}
\label{sec:second_transition_order}

We now determine the order of the transition at $\tau_{c_2}(\phi)$.
We follow the same strategy as in Sec.~\ref{section:Critical Regime}
and show that logarithmic corrections reappear, leading again to a
transition of order $2+0$.
We keep $\phi$ fixed and approach the second critical line from the
two-cut phase, $\tau\to\tau_{c_2}(\phi)^-$. We recall the parametrization
\begin{equation}
    \theta(\tau,\phi)
    =
    \frac{
        \sqrt{4\tau^2e^{2i\phi}+c(\tau,\phi)}
    }{
        \tau e^{i\phi}
    }.
    \label{eq:theta_complex}
\end{equation}
We therefore expand $\theta$ around the one-cut branch as
\begin{equation}
    \theta(\tau,\phi)
    =
    \theta_{\rm 1-cut}(\tau,\phi)+\delta,
    \qquad
    \theta_{\rm 1-cut}(\tau,\phi)
    =
    2-\frac{i e^{-i\phi}}{\tau},
    \qquad
    |\delta|\ll1.
    \label{eq:delta_second_transition}
\end{equation}
The two-cut solution is determined by the two conditions
\eqref{eq:dipole} and \eqref{equ chimique}. We first consider
Eq.~\eqref{eq:dipole}. The elliptic decomposition
\eqref{Decomp_elliptique} remains valid after the substitution
$\tau\to\tau e^{i\phi}$. At the second critical line, the corresponding
elliptic modulus approaches $k\to0$, so that the elliptic integrals
remain regular. Using the small-$k$ expansions collected in
Appendix~\ref{App:Elliptic} and expanding to leading order in $\delta$,
the condition \eqref{eq:dipole} reduces to
\begin{equation}
    \operatorname{Re}\!\left(\beta_c\,\delta\right)=0,
    \qquad
    \beta_c=
    e^{i\phi}
    \sqrt{
        1-\frac{i}{2\tau_{c_2}e^{i\phi}}
    }.
    \label{eq:delta_relation_second}
\end{equation}
Thus the first condition fixes the direction along which
$\theta$ approaches  $\theta_{c_2}(\phi)$ in the complex plane.
This contrasts with the first critical line, where the logarithmic
singularity already appears in Eq.~\eqref{eq:dipole}. The difference follows
from the local topology of the two degenerations. At $\tau_{c_1}(\phi)$,
the two-cut solution emerges from an \textit{ungapped} one-cut configuration, whereas
at $\tau_{c_2}(\phi)$ one of the two cuts becomes empty and collapses into
a \textit{gapped}  one-cut phase. 

The singular behavior instead appears in the second condition, which imposes equality of the electrostatic potentials between the two cuts:
\begin{equation}
    \operatorname{Re}\!\left(
        \int_{z_1}^{z_3} dz\,y(z,c,\phi)
    \right)=0,
    \label{eq:second_equilibrium_period}
\end{equation}
where $z_1$ and $z_3$ are the branch points of the electric field
\eqref{def: y complexe}:
\begin{equation}
\begin{aligned}
z_{1,2}
&=
\frac{1}{2}
\left(
\theta-\frac{i e^{-i\phi}}{\tau}
\pm
\sqrt{
\left(
\theta-\frac{i e^{-i\phi}}{\tau}
\right)^2-4
}
\right),
\\[6pt]
z_{3,4}
&=
\frac{1}{2}
\left(
-\theta-\frac{i e^{-i\phi}}{\tau}
\pm
\sqrt{
\left(
\theta+\frac{i e^{-i\phi}}{\tau}
\right)^2-4
}
\right).
\end{aligned}
\end{equation}
After the change of variable $u=z+z^{-1}$, the integral
in \eqref{eq:second_equilibrium_period} becomes
\begin{align}
    \int_{z_1}^{z_3} dz\,y(z,c,\phi)
    =
    i\tau e^{i\phi}
    \int_{\theta-\frac{ie^{-i\phi}}{\tau}}
         ^{-\theta-\frac{ie^{-i\phi}}{\tau}}
    du\,
    \sqrt{
        \frac{
            \left(
                u-\theta+\frac{ie^{-i\phi}}{\tau}
            \right)
            \left(
                u+\theta+\frac{ie^{-i\phi}}{\tau}
            \right)
        }{
            (u-2)(u+2)
        }
    }.
    \label{eq:B_period_second_transition}
\end{align}
which can also be reduced to a linear combination of elliptic
integrals,
\begin{align}
    &\int_{\theta-\frac{ie^{-i\phi}}{\tau}}
         ^{-\theta-\frac{ie^{-i\phi}}{\tau}}
    du\,
    \sqrt{
        \frac{
            \left(
                u-\theta+\frac{ie^{-i\phi}}{\tau}
            \right)
            \left(
                u+\theta+\frac{ie^{-i\phi}}{\tau}
            \right)
        }{
            (u-2)(u+2)
        }
    }
    =
    \widetilde{A}(\theta,\tau,\phi)E(\widetilde{k})
    +
    \widetilde{B}(\theta,\tau,\phi)K(\widetilde{k})
    +
    \widetilde{C}(\theta,\tau,\phi)\Pi(\widetilde{n},\widetilde{k}),
    \label{eq:elliptic_second_transition}
\end{align}
with
\begin{equation}
\begin{aligned}
    \widetilde{A}(\theta,\tau,\phi)
    &=
    \sqrt{
        \left(
            \theta-\frac{ie^{-i\phi}}{\tau}+2
        \right)
        \left(
            \theta+\frac{ie^{-i\phi}}{\tau}+2
        \right)
    },
    \\
    \widetilde{B}(\theta,\tau,\phi)
    &=
    \left(
        \theta+\frac{ie^{-i\phi}}{\tau}-2
    \right)
    \sqrt{
        \frac{
            \theta-\frac{ie^{-i\phi}}{\tau}+2
        }{
            \theta+\frac{ie^{-i\phi}}{\tau}+2
        }
    },
    \\
    \widetilde{C}(\theta,\tau,\phi)
    &=
    \frac{
        2\frac{ie^{-i\phi}}{\tau}
        \left(
            \theta+\frac{ie^{-i\phi}}{\tau}-2
        \right)
    }{
        \sqrt{
            \left(
                \theta-\frac{ie^{-i\phi}}{\tau}+2
            \right)
            \left(
                \theta+\frac{ie^{-i\phi}}{\tau}+2
            \right)
        }
    }.
\end{aligned}
\label{eq:ABC13_second_transition}
\end{equation}The corresponding elliptic modulus and characteristic are
\begin{equation}
    \widetilde{k}
    =
    \sqrt{
        \frac{
            8\theta
        }{
            \left(
                \theta-\frac{ie^{-i\phi}}{\tau}+2
            \right)
            \left(
                \theta+\frac{ie^{-i\phi}}{\tau}+2
            \right)
        }
    },
    \qquad
    \widetilde{n}
    =
    \frac{
        2\theta
    }{
        \theta-\frac{ie^{-i\phi}}{\tau}+2
    }.
    \label{eq:kn_second_transition}
\end{equation}
As $\delta\to0$, both $\widetilde{k}\to1$ and $\widetilde{n}\to1$, while
\begin{equation}
    \frac{1-\widetilde{k}^2}{1-\widetilde{n}}
    \longrightarrow
    \frac{i}{2\tau_{c_2}e^{i\phi}}.
\end{equation}
The limit is therefore non-uniform, and the expansion of $\Pi(\widetilde{n},\widetilde{k})$
for $\widetilde{k}\to1$ at fixed $\widetilde{n}$ cannot be used. Instead, $\widetilde{k}$ and $\widetilde{n}$ must be
expanded simultaneously, using the asymptotics derived in
Appendix~\ref{App:Elliptic}.

Using these asymptotic expansions in
Eq.~\eqref{eq:elliptic_second_transition}, together with the expansions
of $\widetilde{A}$, ${B}$, and $\widetilde{C}$, and expanding the regular terms to
first order in $\tau-\tau_{c_2}$, the zeroth-order contribution
reproduces the criticality condition   \eqref{eq:second_complex_boundary_explicit}, while the
leading non-analytic correction is proportional to
$\delta\log\delta$. Since Eq.~\eqref{eq:delta_relation_second} fixes
the complex direction of $\delta$, the equilibrium condition therefore
implies
\begin{equation}
    \delta
    \propto
    \frac{
        \tau_{c_2}(\phi)-\tau
    }{
        \log(\tau_{c_2}(\phi)-\tau)
    },
    \qquad
    \tau\to\tau_{c_2}(\phi)^-.
    \label{eq:delta_scaling_second}
\end{equation}
Using Eq.~\eqref{eq:theta_complex}, we finally obtain
\begin{equation}
    c(\tau,\phi)
    =
    -1-4\ii\tau e^{\ii\phi}
    +
    \Gamma(\phi,\tau_{c_2})
    \frac{
        \tau_{c_2}(\phi)-\tau
    }{
        \log\!\left(\tau_{c_2}(\phi)-\tau\right)
    },
    \qquad
    \tau\to\tau_{c_2}(\phi)^-.
    \label{eq:critical_c_scaling_one_cut}
\end{equation}
Combining \eqref{eq:critical_c_scaling_one_cut} with \eqref{lien Setc} 
we obtain, near the second critical line,
\begin{equation}
\label{eq:second_transition_action}
\frac{\partial S}{\partial \tau}
=
\begin{cases}
\displaystyle
\frac{1}{2\tau}
+2\ii e^{\ii\phi}
-\frac{\Gamma(\phi,\tau_{c_2})}{2\tau_{c_2}(\phi)}
\frac{\tau_{c_2}(\phi)-\tau}
{\log\!\left(\tau_{c_2}(\phi)-\tau\right)}
&
\tau\to\tau_{c_2}(\phi)^-,
\\[12pt]
\displaystyle
\frac{1}{2\tau}
+2\ii e^{\ii\phi}
&
\tau\to\tau_{c_2}(\phi)^+ .
\end{cases}
\end{equation}

\begin{figure}[h]
\centering
\includegraphics[width=1.\textwidth]{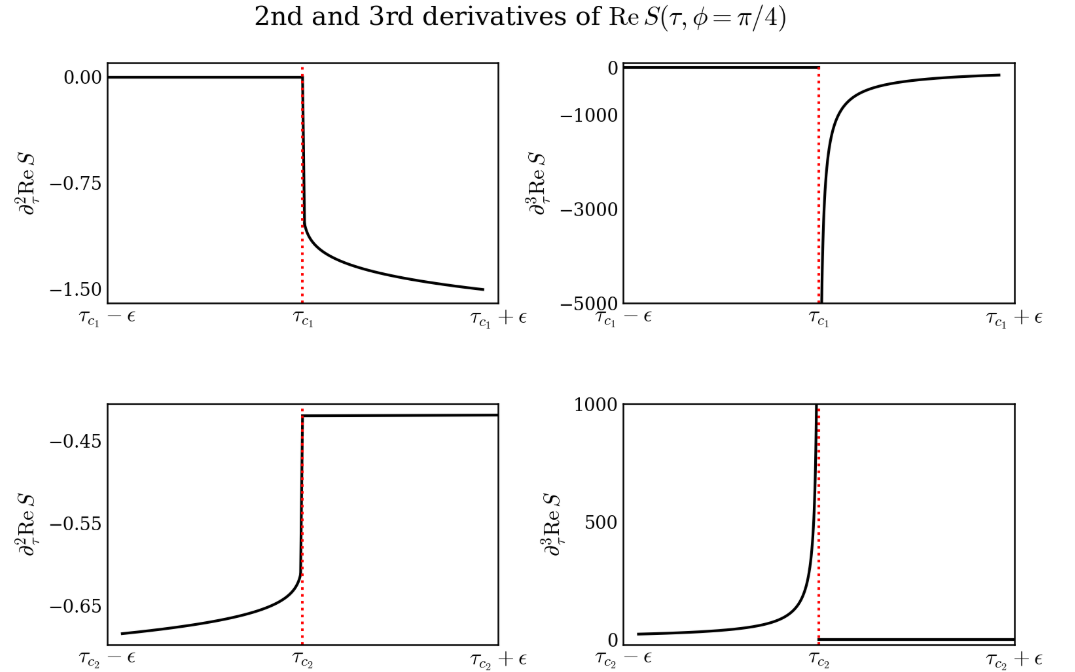}
\caption{
Second and third derivatives of $\operatorname{Re}S(\tau,\phi=\pi/4)$ in the vicinity of the two critical points $\tau_{c_1}$ (top row) and $\tau_{c_2}$ (bottom row), with $\epsilon=10^{-3}$, obtained numerically by solving Eqs.~\eqref{eq:dipole} and~\eqref{eq:second_equilibrium_period}.
}
\label{image: derives action}
\end{figure}
The precise value of the critical constant $\Gamma(\phi,\tau_{c_2})$ is irrelevant for determining the scaling behavior. The logarithmic correction in
Eq.~\eqref{eq:second_transition_action} vanishes as
$(\tau_{c_2}-\tau)/\log|\tau-\tau_{c_2}|$, implying that the second
derivative of $S$ remains continuous, while the third derivative
diverges at $\tau_{c_2}$. The second transition is therefore also of
order $2+0$, in excellent agreement with the numerical behavior shown
in Fig.~\ref{image: derives action} for $\phi=\pi/4$.

\subsubsection{Instanton interpretation}
\label{sec:complex_instanton}

The equation \eqref{eq:second_complex_boundary_explicit} for the second
critical line was also derived in Ref.~\cite{copetti2022delayed} in a different
context. Here, we briefly explain its relation to our electrostatic construction
and its interpretation in terms of an eigenvalue instanton.

Starting from the one-cut configuration, consider transferring an infinitesimal
filling fraction $\delta\nu$ from the occupied cut to the double point $z=-1$.
The corresponding variation of the density can be written as
\begin{equation}
\delta\rho(z)
=
\delta\nu
\left(
\delta\left(z+1\right)-\delta\left(z-a\right)
\right),
\end{equation}
where $a$ is given by \eqref{eq:a_one_cut_complex}.
The corresponding action cost per unit filling fraction defines
the instanton action $S_{\mathrm{inst}}$:
\begin{equation}
\operatorname{Re} S_{\mathrm{inst}}
\simeq
\frac{
\operatorname{Re}S[\rho+\delta\rho]
-
\operatorname{Re}S[\rho]
}{
\delta\nu
}
=
\operatorname{Re}\!\Big(
\Phi_{\mathrm{1-cut}}(-1)
-
\Phi_{\mathrm{1-cut}}(a)
\Big).
\label{eq:instanton_action_complex}
\end{equation}
Our condition ~\eqref{eq:second_complex_boundary} is therefore equivalent to 
\begin{equation}
\operatorname{Re}S_{\mathrm{inst}}=0.
\label{eq:instanton_critical_complex}
\end{equation}
For $\operatorname{Re}S_{\mathrm{inst}}>0$, instanton tunnelling is suppressed
and the one-cut solution remains dominant. When
$\operatorname{Re}S_{\mathrm{inst}}<0$, eigenvalue tunnelling is favored,
signalling an instability of the one-cut solution and the expected dominance
of a two-cut configuration.

\subsection{Phase diagram and Wick rotation}
\label{sect : Phase Diagram}

\begin{figure}[H]
    \centering
    \includegraphics[width=0.8\textwidth]{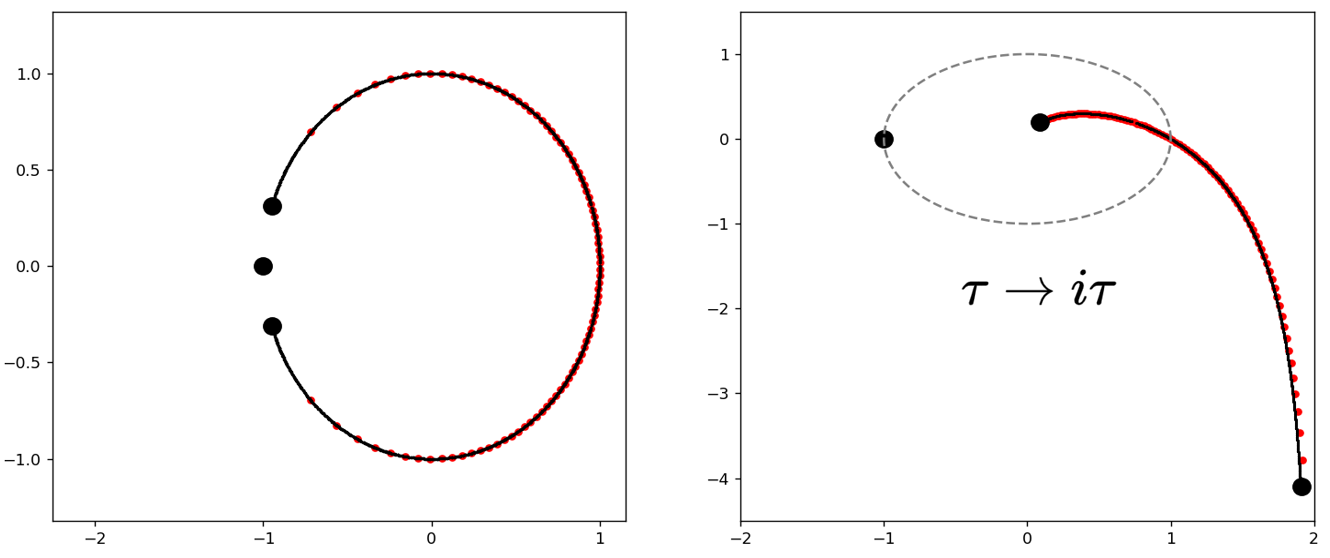}
    \caption{
    Illustration of the failure of a naive Wick rotation at $\tau=0.51$
    and $N=90$. Left: dominant gapped one-cut solution in imaginary time.
    Right: real-time one-cut saddle obtained from the analytically continued
    branch $c_{\mathrm{1-cut}}(\tau,\phi)$ of
    Eq.~\eqref{eq:c_one_cut_complex}. 
    }
    \label{fig:Wick}
\end{figure}

The resulting phase diagram is shown in Fig.~\ref{fig:phase_diagram}.
We restricted ourselves to $t\geq0$ and $\phi\in[0,\pi/2]$, but the remaining
regions follow by symmetry. Three phases are separated by the two
critical lines $\tau_{c_1}(\phi)$ and $\tau_{c_2}(\phi)$.

For $\tau<\tau_{c_1}(\phi)$, the support consists of a single ungapped cut.
Crossing the first critical line produces a second cut, leading to the
intermediate two-cut phase for $\tau_{c_1}(\phi)<\tau<\tau_{c_2}(\phi)$.

At the second critical line, one of the two cuts becomes depleted and
disappears, leaving a gapped one-cut configuration for
$\tau>\tau_{c_2}(\phi)$.
The limiting cases $\phi=0$ and $\phi=\pi/2$ follow naturally from this picture. As $\phi\to0^+$, the second critical line is pushed to $\tau\to\infty$. For exactly $\phi=0$, the evolution never crosses this second boundary, so the two-cut phase persists throughout the long-time regime. Purely imaginary time, $\phi=\pi/2$, is the other special endpoint of the complex-time phase diagram: the two critical lines separating the ungapped one-cut, two-cut, and gapped one-cut phases merge at $\tau_c=1/2$, and the transition reduces to the opening of a gap.

This phase diagram also clarifies the limitation of a naive Wick rotation.
The gapped one-cut saddle obtained in imaginary time can be analytically
continued towards real time, but this continuation does not lead 
to the dominant one. Whenever we enter the intermediate two-cut phase, the dominant saddle instead requires a redistribution of charges between the two components of the support, whereas analytic continuation of the one-cut branch does not produce the required charge transfer.
The correct solution in this phase must therefore be obtained 
by solving
Eqs.~\eqref{eq:dipole} and \eqref{equ chimique}. 

The one-cut solution $c_{\mathrm{1-cut}}(\tau,\phi)$ introduced in
Eq.~\eqref{eq:c_one_cut_complex} can be viewed as the analytic continuation
of the gapped imaginary-time configuration \cite{alvarez_16}.  Evaluating this solution at $\phi=0$ yields a valid
one-cut real-time saddle, shown in Fig.~\ref{fig:Wick}, but not the dominant
one.
Indeed at $\tau=0.51$, the real-time evolution lies in the intermediate two-cut
phase. The dominant saddle is therefore not obtained by analytic continuation
of the one-cut branch, but by solving Eqs.~\eqref{eq:dipole} and
\eqref{equ chimique}. This is also confirmed by
Fig.~\ref{figure dipole}, where the corresponding $N\to\infty$ solution agrees
with the finite-$N$ Toeplitz results throughout the two-cut regime. Thus,
analytic continuation of the one-cut solution across the intermediate phase
does not follow the dominant saddle.

In terms of work statistics, the short-time regime corresponds to a
Gaussian distribution, whereas non-Gaussian tails emerge beyond the
critical line.
For a broader discussion of work statistics in sudden
quantum quenches, we refer the reader to Ref.~\cite{tierz2025}.

\section{Discussion}
\label{sec:disc}

We investigated the Loschmidt echo of non-interacting fermions released
from a DDW initial configuration in the scaling limit $N,t\to\infty$
with $\tau=t/N$ fixed. Rather than relying on an analytic continuation
from imaginary time, we determined directly the dominant complex saddle
configuration governing the real-time dynamics. Our analysis shows that
the DQPT is associated with a topological change of this configuration:
at $\tau=\tau_c$, the connected one-cut support splits into two symmetric
components, in direct analogy with the birth of a cut in random matrix
theory. This change of topology produces a logarithmic correction to the
dynamical free energy and leads to a transition of order $2+0$, distinct
from the third-order GWW transition occurring in imaginary time.

The extension to complex time leads to a richer phase structure.
For generic $0<\phi<\pi/2$, the dominant saddle configuration passes
successively through three topologically distinct phases: an ungapped one-cut
phase, a two-cut phase, and a gapped one-cut phase. The two critical boundaries correspond
respectively to the birth and death of a cut and exhibit the same
logarithmic singularity. Consequently, post-selection can drive the system across both critical boundaries, leading to two successive DQPTs.
This phase diagram
also makes clear why a naive Wick rotation fails: in the intermediate
region, analytic continuation follows a one-cut saddle that is no longer
dominant, whereas the physical saddle requires a redistribution of
charges between the two components of the support. In this sense, our
results complete the picture of the
complex-coupling GWW model discussed in Ref.~\cite{copetti2022delayed}
by resolving the structure of the intermediate multi-cut phase and its
boundaries.
An interesting future direction would be to explore whether
the DQPT persists for more general initial states. The DDW state is very special and atypical: it has zero entropy and no initial coherence or fluctuations between sites. One may consider smoother initial states to probe the influence of initial fluctuations on the existence and nature of the DQPT. 
\begin{figure}[H]
\begin{center}
\includegraphics[angle=-90,width=.475\linewidth]{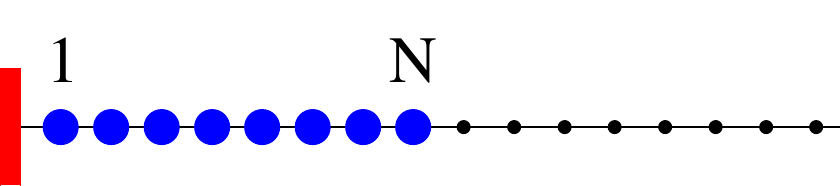}
\caption{Initial configuration of $N$ fermions near a wall
ending a semi-infinite chain.}
\label{fig:wall}
\end{center}
\end{figure}
One can consider the Loschmidt echo for non-interacting fermions on substrates different from the one-dimensional lattice. In one dimension, it would be interesting to explore the nature of the DQPT in a semi-infinite lattice chain ending with an impenetrable wall. Starting with the extreme configuration (Fig.~\ref{fig:wall}), the evolution of the return probability was studied in \cite{KLM18}. In the $t\to\infty$ limit with $N$ kept finite, the return probability decays algebraically with exponent $N^2 + N$ for even $N$ and $N^2 + N+1$ for odd $N$, where the power law decay is modulated by periodic oscillations. The scaling behavior \eqref{LN:scaling} continues to hold, and the asymptotic behaviors of $S(\tau)$ for $\tau\ll 1$ and $\tau\gg 1$ are the same \cite{KLM18} as in the DDW case. Conjecturally, there is again a DQPT of order $2+0$, the behavior in the $\tau<\tau_c$ regime is Gaussian, and only the value of $\tau_c$ and the details of the behavior in the $\tau>\tau_c$ regime quantitatively differ from the DQPT in the  DDW case. The imaginary-time evolution of the configuration shown in Fig.~\ref{fig:wall} was studied in \cite{Page_chain}, where it was argued that it mimics the Hawking-Page transition in a supersymmetric model. 

Studying the Loschmidt echo for non-interacting fermions in higher dimensions is an interesting avenue for future work. Even for maximally dense initial configurations, the shape of the initial state is expected to play a role. For instance, for the fermions on the square grid, the dynamical free energy probably depends on whether the fermions were released from the square or rhombus, although the nature of the DQPT may be universal. 

A natural next step is to study interacting fermions. Some work in this direction has already been done, e.g., in the realm of the XXZ spin chain, where the Loschmidt echo for the DW initial configuration,
$\ket{\Psi_0}=\ket{\cdots \uparrow \uparrow \uparrow \downarrow \downarrow \downarrow \cdots}$,
has been studied both in real and imaginary time \cite{stephan_17,stephan_22}.

It would also be interesting to identify observables that could directly signal the DQPT and, hopefully, be more analytically tractable than the Loschmidt echo.

\paragraph{Acknowledgements}
K.M. and S.P. are grateful  to Fran\c{c}ois David and Bertrand  Eynard for illuminating discussions on loop equations in the complex plane.


\paragraph{Funding information}
 The work of KM is supported by the project RETENU ANR-20-CE40-0005-01 of the French National Research Agency (ANR). D.M.G. acknowledges support from the Leverhulme Trust under Grant
No.~RPG-2024-401.

\begin{appendix}
\numberwithin{equation}{section}

\section{Analysis from the Toeplitz determinant representation}
\label{Appendix:Toeplitz}

In this appendix, we show how the Toeplitz determinant representation
\niceref{eq:toeplitz} can be used to obtain the short-time behavior and
the existence of a critical value $\tau_c$.

\subsection{Short-time asymptotics}
\label{Appendix : Scego}

The Gaussian decay of the Loschmidt echo in the short-time regime described by Eq.~\eqref{eq:LE_short} follows from the strong Szeg\"o theorem governing the asymptotic behavior of large Toeplitz determinants \cite{Bottcher}:  if the coefficients of a Toeplitz matrix  $T^{(N)}$ of size $N$ are given by
\begin{equation}
T^{(N)}_{n,m} = \hat\phi(n-m) \, , \quad \hat\phi(j) = \frac{1}{2\pi}\int_{-\pi}^{\pi} \phi(\theta)
   {\rm e}^{-\ii j \theta} {\rm d}\theta \, , 
\end{equation}
and $\phi$ is a smooth enough function (twice differentiable suffices), then
\begin{equation}
\lim_{N \to \infty} {\rm e}^{-N \, \widehat{\log\phi}(0)} \det T^{(N)}
   = \exp\!\left(\sum_{k\geq 1} k \, \, \widehat{\log\phi}(k) \,\, \widehat{\log\phi}(-k) \right).
\end{equation}
Here, we have a  Toeplitz determinant of Bessel functions and thus   $\log\phi(\theta) = 2 \ii t \cos\theta$, which implies 
$\widehat{\log\phi}(k) = \ii t (\delta_{k,1} + \delta_{k,-1}) $  and the asymptotics (\ref{eq:LE_short}) follows.

\subsection{Alternative derivation of the DQPT using exact recursion relation}
\label{Appendix Recursion}
An exact recursion relation can be derived for the Toeplitz determinant~\eqref{eq:toeplitz}, which is very useful for numerical analysis, allowing us to compute the Loschmidt amplitude for very large values of $N$ with high precision.
 Let us first define the quantity $w_N$ 
\begin{equation}
  w_N^2(t) = (-1)^{N}\bigg(1 - \frac{r_{N+1}(t)}{r_N(t)} \, \bigg) \, ,
\label{def:WN}
 \end{equation}
where we defined the ratio $r_N(t)=\frac{A_N(t)}{A_{N-1}(t)}$. 
This leads to 
\begin{equation}
    r_{N+1}=r_N\Big(1-(-1)^Nw_N^2\Big) \, .
    \label{suite rn}
\end{equation}
One can then show that $w_N$ satisfies
 the following non-linear recursion relation, related 
 to the discrete Painlev\'e II equation  \cite{borodin2003,adler2001,baik2001}:
\begin{equation}
  w_{N -1} -  w_{N +1} = \frac{N}{t} \frac{w_N}{1 - (-1)^N w_N^2} \, ,
  \label{Recursion}
 \end{equation}
with $w_0 =1$ and  $w_1 =J_1(2t)/J_0(2t)$.
Thus, one can  obtain $w_n$ and $r_n$  by recursion  and deduce $A_N(t)$ as
\begin{equation}
    A_N(t)=A_0(t)\prod_{k=1}^Nr_k(t) \, .
\end{equation}
Moreover, the recursion relation \eqref{Recursion} provides a convenient way to identify heuristically the critical value $\tau_c$ at which the DQPT occurs.
Let us assume that there exists  a regime such that $w_N(t) \ll 1$. In this regime, \eqref{Recursion} can be linearized and reduces to 
\begin{equation}
  w_{N-1} - w_{N+1} = \frac{N}{t}\, w_N .
  \label{linear_recursion}
\end{equation}
The solution of this linearized recursion is given by the imaginary Bessel functions
$$ w_N \simeq I_N(2t) = \sum_{k \ge 0} \frac{t^{N+2k}}{k! (N+k)!}  \, .$$ 
The asymptotic behavior of the  imaginary Bessel function for large $N$ and $t$ is (see \cite{Abramowitz1964}) 
\begin{equation}
  I_N(2t) \simeq {\rm e}^{ 2t (\cosh \eta - \eta \sinh \eta)} \, , 
  \end{equation}
where we have introduced the parameter $\eta$ defined as
$2\sinh\eta = \frac{1}{\tau}$. 
The assumption $w_N \ll 1$ is consistent as long as
$\cosh\eta - \eta\sinh\eta \le 0$.
It breaks down at the critical value $\eta_c$
satisfying
\begin{equation}
  \eta_c \tanh\eta_c = 1  \, .
  \label{criticJML}
\end{equation}
Eq.~(\ref{criticJML}) is equivalent to Eq.~(\ref{eq: lbd critique}), as it should be.
Beyond this critical point $\eta_c$, the recursion relation cannot be considered as linear anymore and this corresponds to the transition in $A_N(t)$.
This criterion  gives the numerical values
  \begin{eqnarray}
 \eta_c  &=&  1.199678640257  \ldots   \\
 \tau_c = \frac{1}{2 \sinh\eta_c}  &=&  0.331 371 709 674  \ldots
 \label{LambdacNum}
 \end{eqnarray}

\section{Electrostatic interpretation and Riemann-Hilbert problem}
\label{App:Electrostatic}

The Coulomb gas method is a well-established tool for studying the leading large-$N$ asymptotics of matrix models. It can be viewed as an effective continuous description of the saddle-point approximation. In our context, this approach allows us to replace the problem of finding the density that dominates the path integral by an equivalent electrostatic equilibrium problem.  Indeed, the real part of the action \eqref{eq:action} can be interpreted as the total electrostatic energy of a charge density $\rho(z)$ supported on the contour $\gamma$. These $N$ charges experience an external potential $V(z)$ and interact through a repulsive logarithmic Coulomb potential. Likewise, the real part of  $\Phi(z)$ can be interpreted as the total electrostatic potential generated by this configuration.
We can then define a complex electric field using  the derivative of the electrostatic potential:
\begin{equation}
	E(z) = -\frac{d\Phi(z)}{d z} = -\overline{y(z)} \, .
\end{equation}
Using the definition of the electric field  $(\ref{eq:spectral_curve})$ and the Sokhotski--Plemelj formula,  the electric field  can be expressed
as a sum of the limiting values, 
\begin{equation}
    y(z) = \frac{y_+(z)+y_-(z)}{2} \quad\text{for }z \in \gamma \, , 
\end{equation}
which allows us  to express the condition (\ref{Anti stockes line}) as an electrostatic
equilibrium condition along the contour $\gamma$:
\begin{equation}
    y_+(z)+y_-(z)=0 \quad\text{for }z \in \gamma \, .
    \label{equ: equilibrium electric}
\end{equation}
\begin{figure}[H]
    \centering
    \includegraphics[width=0.4\textwidth]{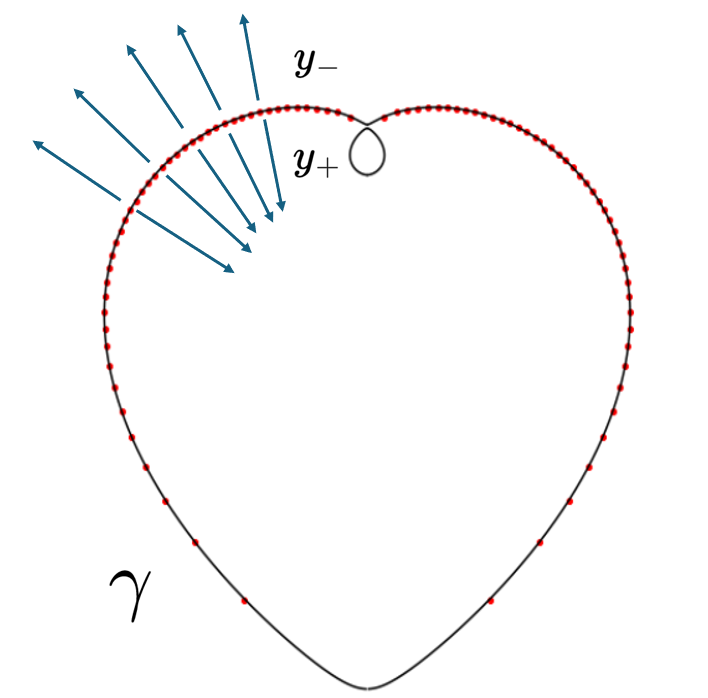}
\caption{Illustration of the electrostatic equilibrium on $\gamma$.}
\end{figure}
Alternatively, by using ~\eqref{eq:spectral_curve}, the electrostatic equilibrium condition can be reformulated as a
Riemann-Hilbert problem:
\begin{equation}
    \omega_+(z)+\omega_-(z)=V'(z)\quad\text{for }z \in \gamma \, .
    \label{eq:Riemann-Hilbert}
\end{equation}
In standard Riemann-Hilbert problems, the potential is usually real, and the jump contour of the resolvent $\omega(z)$ lies on an arc of the unit circle, as in \cite{GrossWitten1980,WADIA}. In this case, the problem reduces to determining a positive density satisfying the corresponding jump condition. Here, however, the situation is more involved: the potential $V(z)$ is complex-valued and, as explained in the main text, the contour $\gamma$ is no longer fixed but must be determined as part of the solution.

\section{Derivation of the loop equation}
\label{App:loop}
We recall here the finite-$N$ derivation of the loop equation before taking its thermodynamic limit. Rewriting the saddle-point condition \eqref{eq:saddle point q} in terms of $z_i=e^{iq_i}$ gives
\begin{equation}
\label{eq:discrete saddle point}
V'(z_i)
-
\frac{2}{N}\sum_{j\neq i}\frac{1}{z_i-z_j}
=
0,
\end{equation}
and we introduce the discrete resolvent
\begin{equation}
\omega_N(z)
=
\frac{1}{N}\sum_{i=1}^N\frac{1}{z-z_i}.
\end{equation}
Multiplying Eq.~\eqref{eq:discrete saddle point} by
$\frac{1}{N(z-z_i)}$ and summing over $i$ yields
\begin{equation}
   \frac{1}{ N} \sum_{i=1}^N \frac{V'(z_i)}{z - z_i}  -\frac{2}{N^2} \sum_{i=1}^N\sum_{j \neq i}^N \frac{1}{(z - z_i)(z_i - z_j)}  = 0 \, .
   \label{equ: Loop non simplif}
\end{equation}
The first term becomes
\begin{equation}
   \frac{1}{ N} \sum_{i=1}^N \frac{V'(z_i)}{z - z_i}  =V'(z)\omega_N(z)-\frac{1}{ N}\sum_{i=1}^N \frac{V'(z)-V'(z_i)}{z - z_i} \, . 
\end{equation}
Symmetrizing the second term and using the definition of the resolvent gives
\begin{align}
    \frac{2}{N^2}\sum_{i=1}^N\sum_{j \neq i}^N \frac{1}{(z - z_i)(z_i - z_j)}
    &= \frac{1}{N^2}\sum_{i=1}^N\sum_{j\neq i}^N \left( \frac{1}{(z - z_i)(z_i - z_j)} + \frac{1}{(z - z_j)(z_j - z_i)} \right) \nonumber \\
    &= \frac{1}{N^2} \sum_{i=1}^N\sum_{j\neq i}^N  \frac{1}{(z - z_i)(z - z_j)}\nonumber\\
    &= \bigg(\frac{1}{N} \sum_{i=1}^N\frac{1}{(z - z_i)}\bigg)^2-\frac{1}{N^2}\sum_{i=1}^N\frac{1}{(z-z_i)^2}\nonumber\\
    &=\omega_N^2(z)+\frac{1}{N}\omega_N'(z) \, .
\end{align}
Substituting these expressions into Eq.~\eqref{equ: Loop non simplif} gives the loop equation for the resolvent,
\begin{equation}
    V'(z)\omega_N(z)-\omega_N^2(z) - \frac{1}{N} \omega_N'(z) 
    =  \frac{1}{N} \sum_i \frac{V'(z) - V'(z_i)}{z - z_i} \, , 
    \label{equ: Loop discrete}
\end{equation}
which reduces to \eqref{equ: Loop equation} in the thermodynamic limit, thereby replacing the $N$ coupled nonlinear saddle point equations by a single quadratic equation for the resolvent.

\section{Elliptic integrals and their expansions}
\label{App:Elliptic}
In this Appendix, we collect the asymptotic expansions of the complete
elliptic integrals $K(k)$, $E(k)$, and $\Pi(n,k)$ used throughout the
paper, following Refs.~\cite{Byrd,wolfram_elliptic}.
 
We first recall the three canonical forms of the complete elliptic integrals 
\begin{align}
K(k)
&=
\int_0^{\pi/2}
\frac{d\theta}{\sqrt{1-k^2\sin^2\theta}},
\\
E(k)
&=
\int_0^{\pi/2}
d\theta\,\sqrt{1-k^2\sin^2\theta},
\\
\Pi(n,k)
&=
\int_0^{\pi/2}
\frac{d\theta}
{(1-n\sin^2\theta)\sqrt{1-k^2\sin^2\theta}}.
\end{align}

\subsection{Asymptotics for \texorpdfstring{$k\to1$}{k to 1}}

\label{Appendix : asymptos k1}
\begin{align}
K(k)
\simeq&\,
\log\frac{4}{\sqrt{1-k^2}}
+
\frac{1-k^2}{4}
\left(
\log\frac{4}{\sqrt{1-k^2}}-1
\right)+
\frac{9(1-k^2)^2}{64}
\left(
\log\frac{4}{\sqrt{1-k^2}}-\frac{7}{6}
\right)
\\[2mm]
E(k)\simeq&\,
1
+
\frac{1-k^2}{2}
\left(
\log\frac{4}{\sqrt{1-k^2}}-\frac{1}{2}
\right)+
\frac{3(1-k^2)^2}{16}
\left(
\log\frac{4}{\sqrt{1-k^2}}-\frac{13}{12}
\right)
\end{align}
For fixed $n\neq1$:
\begin{align}
\Pi(n,k)\simeq &\,
\frac{1}{1-n}
\left(
\log\frac{4}{\sqrt{1-k^2}}
-
\sqrt{n}\,\operatorname{artanh}\sqrt{n}
\right)+
\frac{1-k^2}{4(1-n)^2}
\left(
(1+n)\log\frac{4}{\sqrt{1-k^2}}
-
2\sqrt{n}\,\operatorname{artanh}\sqrt{n}
-1
\right)
\nonumber\\
&+
\frac{(1-k^2)^2}{128(1-n)^3}
\Bigg(
6(3+6n-n^2)
\log\frac{4}{\sqrt{1-k^2}}
+
5n^2-12n-21
-
48\sqrt{n}\,\operatorname{artanh}\sqrt{n}
\Bigg)
\end{align}

\subsection{Asymptotics for \texorpdfstring{$k\to0$}{k to 0}}
\begin{align}
K(k)
&\simeq
\frac{\pi}{2}
\left(
1+\frac{k^2}{4}+\frac{9k^4}{64}
\right),
\\[0.5em]
E(k)
&\simeq
\frac{\pi}{2}
\left(
1-\frac{k^2}{4}-\frac{3k^4}{64}
\right).
\end{align}

For fixed $n\neq1$,
\begin{equation}
\Pi(n,k)
\simeq
\frac{\pi}{2\sqrt{1-n}}
+
\frac{\pi}{4\sqrt{1-n}\left(1+\sqrt{1-n}\right)}\,k^2
+
\frac{3\pi\left(2+\sqrt{1-n}\right)}
{32\sqrt{1-n}\left(1+\sqrt{1-n}\right)^2}\,k^4.
\end{equation}
\subsection{Asymptotics of \texorpdfstring{$\Pi(n,k)$}{Pi(n,k)} for both
\texorpdfstring{$k\to1$ and $n\to1$}{kto1 and nto1}}

The complete elliptic integral of the third kind can be expressed as
\begin{equation}
    \Pi(n,k)
    =
    K(k)
    +
    \frac{n}{2}
    \int_0^\infty
    \frac{dt}{
        (t+1-n)\sqrt{t(t+1-k^2)(t+1)}
    }.
\end{equation}
The asymptotic behavior of $K(k)$ in this limit was given previously
in Sec.~\ref{Appendix : asymptos k1}. It therefore remains to determine
the asymptotics of the second term. To this end, we introduce
\begin{equation}
    \Delta_k=1-k^2,
    \qquad
    \Delta_n=1-n,
    \qquad
    r=\frac{\Delta_k}{\Delta_n}.
\end{equation}
In the vicinity of the second critical line discussed in
Sec.~\ref{sec:second_transition_order}, both $\Delta_k$ and $\Delta_n$
vanish linearly in $\delta=\theta-\theta_{c_2}$ and are therefore of the
same order.

We first split the integral into two regions:
\begin{align}
\int_0^\infty
\frac{dt}{
(t+\Delta_n)\sqrt{t(t+\Delta_k)(t+1)}
}
=
&\int_0^\lambda
\frac{dt}{
(t+\Delta_n)\sqrt{t(t+\Delta_k)(t+1)}
}+
\int_\lambda^\infty
\frac{dt}{
(t+\Delta_n)\sqrt{t(t+\Delta_k)(t+1)}
}.
\end{align}
Here, $\lambda$ is an intermediate cutoff satisfying
$|\Delta_k|,|\Delta_n|\ll\lambda\ll1$, chosen such that the integrand
can be expanded consistently in both regions.

In the inner region, $t$ can be of the same order as $\Delta_k$ and
$\Delta_n$. We therefore zoom into this region by performing the change
of variable $t=\Delta_n x$. Expanding for $|\Delta_n x|\ll1$, and integrating leads to
\begin{equation}
\int_0^\lambda
\frac{dt}{
(t+\Delta_n)\sqrt{t(t+\Delta_k)(t+1)}
}
\simeq
\int_0^{\lambda/\Delta_n}
\frac{dx}{(1+x)\sqrt{x(x+r)}}
\left(
\frac{1}{\Delta_n}
-\frac{x}{2}
+\frac{3\Delta_n x^2}{8}
\right).
\end{equation}

Evaluating the resulting integrals gives
\begin{equation}
\begin{aligned}
\int_0^\lambda
\frac{dt}{
(t+\Delta_n)\sqrt{t(t+\Delta_k)(t+1)}
}
\simeq{}&
\frac{
2/\Delta_n+1+3\Delta_n/4
}{
\sqrt{r-1}
}
\arctan\!
\sqrt{
\frac{(r-1)\lambda}
{\lambda+\Delta_k}
}
\\
&-
\left(
1+\frac{3}{8}(\Delta_k+2\Delta_n)
\right)
\operatorname{arsinh}\!
\sqrt{\frac{\lambda}{\Delta_k}}
+
\frac{3}{8}\sqrt{\lambda(\lambda+\Delta_k)} .
\end{aligned}
\label{eq:RJ_inner}
\end{equation}
In the outer region, $t\geq\lambda$, so that both
$|\Delta_k|/t$ and $|\Delta_n|/t$ are small. Expanding to second order and then integrating, we obtain 
\begin{equation}
\begin{aligned}
&\int_\lambda^\infty
\frac{dt}{
(t+\Delta_n)\sqrt{t(t+\Delta_k)(t+1)}
}
\\
&\qquad\simeq
\frac{\sqrt{1+\lambda}}{\lambda}
-
\operatorname{artanh}\!
\left(
\frac{1}{\sqrt{1+\lambda}}
\right)
-\left(
\Delta_n+\frac{\Delta_k}{2}
\right)
\left(
\frac{\sqrt{1+\lambda}(2-3\lambda)}
{4\lambda^2}
+
\frac{3}{4}
\operatorname{artanh}\!
\left(
\frac{1}{\sqrt{1+\lambda}}
\right)
\right)
\\
&\qquad\quad
+\left(
\Delta_n^2
+\frac{\Delta_k\Delta_n}{2}
+\frac{3\Delta_k^2}{8}
\right)
\left(
\frac{\sqrt{1+\lambda}
\left(8-10\lambda+15\lambda^2\right)}
{24\lambda^3}
-
\frac{5}{8}
\operatorname{artanh}\!
\left(
\frac{1}{\sqrt{1+\lambda}}
\right)
\right).
\end{aligned}
\label{eq:RJ_outer}
\end{equation}
Finally, by expanding Eqs.~\eqref{eq:RJ_inner} and \eqref{eq:RJ_outer}
for $|\Delta_k|,|\Delta_n|\ll\lambda$ and $\lambda\ll1$, and adding the
two contributions, all  dependence on the  cutoff $\lambda$ cancels.
We therefore obtain
\begin{align}
\int_0^\infty
\frac{dt}{(t+\Delta_n)\sqrt{t(t+\Delta_k)(t+1)}}
&\simeq
\frac{2/\Delta_n+1+3\Delta_n/4}{\sqrt{r-1}}
\arctan\!\sqrt{r-1}
+\frac{1}{2}
+\frac{1}{2}\log\!\left(\frac{\Delta_k}{16}\right)
\nonumber\\
&\quad
+\frac{13\Delta_k+14\Delta_n}{32}
+\frac{3(\Delta_k+2\Delta_n)}{16}
\log\!\left(\frac{\Delta_k}{16}\right)
\label{eq:Pi_integral_second_asymptotic}
\end{align}

\end{appendix}

\bibliography{references_DDW}

@article{adler2001,
	author = {Adler, M and van Moerbeke, P},
	title = {Integrals over classical groups, random permutations, {Toda and Toeplitz} lattices},
	journal = {Comm. Pure Appl. Math.},
	volume = {54},
	pages = {153--205},
	year = {2001},
	url = {https://doi.org/10.1002/1097-0312(200102)54:2<153::AID-CPA2>3.0.CO;2-5},
	doi = {10.1002/1097-0312(200102)54:2<153::AID-CPA2>3.0.CO;2-5}
}

@article{Allegra2016,
	title = {Inhomogeneous field theory inside the arctic circle},
	volume = {2016},
	doi = {10.1088/1742-5468/2016/05/053108},
	number = {5},
	journal = {J. Stat. Mech.},
	author = {Allegra, Nicolas and Dubail, Jérôme and Stéphan, Jean Marie and Viti, Jacopo},
	year = {2016},
}

@article{Altman2021,
  author  = {Altman, Ehud and Brown, Kenneth R. and Carleo, Giuseppe and Carr, Lincoln D. 
             and Demler, Eugene and Eckardt, Andr{\'e} and Foster, Matthew S. 
             and Gorshkov, Alexey V. and Greene, Brian and Hazzard, Kaden R. A. 
             and Rey, Ana Maria and Saffman, Mark and Schleier-Smith, Monika H. 
             and Schollw{\"o}ck, Ulrich and Vuleti{\'c}, Vladan and Lukin, Mikhail D.},
  title   = {Quantum Simulators: Architectures and Opportunities},
  journal = {PRX Quantum},
  volume  = {2},
  number  = {1},
  pages   = {017003},
  year    = {2021},
  doi     = {10.1103/PRXQuantum.2.017003}
}

@article{alvarez_16,
  title = {Complex Saddles in the {{Gross-Witten-Wadia}} Matrix Model},
  author = {Álvarez, Gabriel and Alonso, Luis Martínez and Medina, Elena},
  year = {2016},
  journal = {Phys. Rev. D},
  volume = {94},
  number = {10},
  pages = {105010},
  doi = {10.1103/PhysRevD.94.105010}
}

@article{borodin2003,
	author = {Borodin, A},
        journal = {Duke Math. J.},
	title = {Discrete gap probabilities and discrete {P}ainlevé equations},
	volume = {117},
        pages = {489--542},
	year = {2003},
	doi = {10.1215/S0012-7094-03-11734-2}
}

@article{claeysBirthCutUnitary2008,
  title = {Birth of a {{Cut}} in {{Unitary Random Matrix Ensembles}}},
  author = {Claeys, T.},
  journal = {Int. Math. Res. Notices},
  volume = {2008},
  year = {2008},
  pages = {rnm166},
  doi = {10.1093/imrn/rnm166},
  url = {https://doi.org/10.1093/imrn/rnm166},
}

@article{eynardUniversalDistributionRandom2006,
  title = {Universal Distribution of Random Matrix Eigenvalues near the "Birth of a Cut" Transition},
  author = {Eynard, Bertrand},
  journal = {J. Stat. Mech.},
  volume = {2006},
  number = {07},
  pages = {P07005-P07005},
  year = {2006},
  doi = {10.1088/1742-5468/2006/07/P07005}
}

@article{eynard_random_2018,
	author = {Eynard, Bertrand and Kimura, Taro and Ribault, Sylvain},
	title = {Random matrices},
	journal = {arXiv:1510.04430},
	url = {http://arxiv.org/abs/1510.04430},
	doi = {10.48550/arXiv.1510.04430},
	year = {2018},
}

@article{Eisert2015,
  author  = {Eisert, Jens and Friesdorf, Martin and Gogolin, Christian},
  title   = {Quantum many-body systems out of equilibrium},
  journal = {Nature Physics},
  volume  = {11},
  number  = {2},
  pages   = {124--130},
  year    = {2015},
  doi     = {10.1038/nphys3215}
}

@article{Bloch2012,
  author  = {Bloch, Immanuel and Dalibard, Jean and Nascimb{\`e}ne, Sylvain},
  title   = {Quantum simulations with ultracold quantum gases},
  journal = {Nature Physics},
  volume  = {8},
  number  = {4},
  pages   = {267--276},
  year    = {2012},
  doi     = {10.1038/nphys2259}
}

@article{flaschner_observation_2018,
	title = {Observation of dynamical vortices after quenches in a system with topology},
	volume = {14},
	url = {https://www.nature.com/articles/s41567-017-0013-8},
	doi = {10.1038/s41567-017-0013-8},
	number = {3},
	journal = {Nature Phys},
	author = {Fläschner, N. and Vogel, D. and Tarnowski, M. and Rem, B. S. and Lühmann, D.-S. and Heyl, M. 
	and Budich, J. C. and Mathey, L. and Sengstock, K. and Weitenberg, C.},
	year = {2018},
	pages = {265--268},
}

@article{Gross2017,
  author  = {Gross, Christian and Bloch, Immanuel},
  title   = {Quantum simulations with ultracold atoms in optical lattices},
  journal = {Science},
  volume  = {357},
  number  = {6355},
  pages   = {995--1001},
  year    = {2017},
  doi     = {10.1126/science.aal3837}
}

@article{GrossWitten1980,
	title = {Possible third-order phase transition in the large-{N} lattice gauge theory},
	volume = {21},
	url = {https://link.aps.org/doi/10.1103/PhysRevD.21.446},
	doi = {10.1103/PhysRevD.21.446},
	number = {2},
	journal = {Phys. Rev. D},
	author = {Gross, David J. and Witten, Edward},
	year = {1980},
	pages = {446--453},
}

@article{Heyl2013,
  author  = {Heyl, Markus and Polkovnikov, Anatoli and Kehrein, Stefan},
  title   = {Dynamical Quantum Phase Transitions in the Transverse-Field {I}sing Model},
  journal = {Phys. Rev. Lett.},
  volume  = {110},
  number  = {13},
  pages   = {135704},
  year    = {2013},
  doi     = {10.1103/PhysRevLett.110.135704}
}

@article{Heyl2018,
  author  = {Heyl, Markus},
  title   = {Dynamical Quantum Phase Transitions: A Review},
  journal = {Rep. Prog. Phys.},
  volume  = {81},
  number  = {5},
  pages   = {054001},
  year    = {2018},
  doi     = {10.1088/1361-6633/aaaf9a}
}

@article{Huse_SG,
  title = {Zero-temperature critical behavior of the infinite-range quantum {Ising} spin glass},
  author = {Miller, Jonathan and Huse, David A.},
  journal = {Phys. Rev. Lett.},
  volume = {70},
  issue = {20},
  pages = {3147--3150},
  numpages = {0},
  year = {1993},
  doi = {10.1103/PhysRevLett.70.3147},
  url = {https://link.aps.org/doi/10.1103/PhysRevLett.70.3147}
}

@article{Read_SG,
  title = {Solvable spin glass of quantum rotors},
  author = {Ye, J. and Sachdev, S. and Read, N.},
  journal = {Phys. Rev. Lett.},
  volume = {70},
  issue = {25},
  pages = {4011--4014},
  numpages = {0},
  year = {1993},
  doi = {10.1103/PhysRevLett.70.4011},
  url = {https://link.aps.org/doi/10.1103/PhysRevLett.70.4011}
}

@article{Jurcevic2017,
  author  = {Jurcevic, Petar and Shen, Haifeng and Hauke, Philipp and Maier, Christopher and Hempel, Cornelius and Lanyon, Ben P. and Blatt, Rainer and Roos, Christian F.},
  title   = {Direct Observation of Dynamical Quantum Phase Transitions in an Interacting Many-Body System},
  journal = {Phys. Rev. Lett.},
  volume  = {119},
  number  = {8},
  pages   = {080501},
  year    = {2017},
  doi     = {10.1103/PhysRevLett.119.080501}
}

@article{Touchette2009,
  author  = {Touchette, Hugo},
  title   = {The large deviation approach to statistical mechanics},
  journal = {Phys. Reports},
  volume  = {478},
  pages   = {1--69},
  year    = {2009},
  doi     = {10.1016/j.physrep.2009.05.002}
}

@article{guoObservationDynamicalQuantum2019,
    author = {Guo, X.-Y. and others},
    title = {Observation of a Dynamical Quantum Phase Transition by a Superconducting Qubit Simulation},
    journal = {Phys. Rev. Appl.},
    year = {2019},
    volume = {11},
    pages = {044080},
    DOI = {10.1103/PhysRevApplied.11.044080}    
}

@article{Preskill2018,
  author  = {Preskill, John},
  title   = {Quantum Computing in the {NISQ} Era and Beyond},
  journal = {Quantum},
  volume  = {2},
  pages   = {79},
  year    = {2018},
  doi     = {10.22331/q-2018-08-06-79}
}

@article{Polkovnikov2011,
  author  = {Polkovnikov, Anatoli and Sengupta, Krishnendu and Silva, Alessandro and Vengalattore, Mukund},
  title   = {Colloquium: Nonequilibrium dynamics of closed interacting quantum systems},
  journal = {Rev. Mod. Phys.},
  volume  = {83},
  number  = {3},
  pages   = {863--883},
  year    = {2011},
  doi     = {10.1103/RevModPhys.83.863}
}

@article{KLM18,
        title = {Quantum return probability  of a system of \textit{{N}} non-interacting lattice fermions},
	journal = {J. Stat. Mech.},
	author = {Krapivsky, P L and Luck, J M and Mallick, K},
	year = {2018},
	pages = {P023104}
}

@article{pallister_limit_2022,
	title = {Limit shape phase transitions: a merger of arctic circles},
	volume = {55},
	url = {https://dx.doi.org/10.1088/1751-8121/ac79ad},
	doi = {10.1088/1751-8121/ac79ad},
	number = {30},
	journal = {J. Phys. A: Math. Theor.},
	author = {Pallister, James S. and Gangardt, Dimitri M. and Abanov, Alexander G.},
	year = {2022},
	pages = {304001},
}

@article{stephan_22,
doi = {10.1088/1751-8121/ac5fe8},
url = {https://doi.org/10.1088/1751-8121/ac5fe8},
year = {2022},
volume = {55},
number = {20},
pages = {204003},
author = {Stéphan, Jean-Marie},
title = {Exact time evolution formulae in the {XXZ} spin chain with domain wall initial state},
journal = {Journal of Physics A: Mathematical and Theoretical},
}

@article{Page_chain,
  title = {{Hawking-Page} transition on a spin chain},
  author = {P\'erez-Garc\'{\i}a, David and Santilli, Leonardo and Tierz, Miguel},
  journal = {Phys. Rev. Res.},
  volume = {6},
  issue = {3},
  pages = {033007},
  numpages = {15},
  year = {2024},
  doi = {10.1103/PhysRevResearch.6.033007},
  url = {https://link.aps.org/doi/10.1103/PhysRevResearch.6.033007}
}

@article{stephan_extreme_2021,
	title = {Extreme boundary conditions and random tilings},
	doi = {10.21468/SciPostPhysLectNotes.26},
	journal = {SciPost Phys. Lect. Notes},
	author = {Stéphan, Jean-Marie},
	year = {2021},
	pages = {26},
}

@article{stephan_17,
doi = {10.1088/1742-5468/aa8c19},
url = {https://doi.org/10.1088/1742-5468/aa8c19},
year = {2017},
volume = {2017},
number = {10},
pages = {103108},
author = {Stéphan, Jean-Marie},
title = {Return probability after a quench from a domain wall initial state in the spin-1/2 {XXZ} chain},
journal = {Journal of Statistical Mechanics: Theory and Experiment},
}

@article{moRiemannHilbertApproach2008,
  title = {The {{Riemann}}–{{Hilbert Approach}} to {{Double Scaling Limit}} of {{Random Matrix Eigenvalues Near}} the “{{Birth}} of a {{Cut}}” {{Transition}}},
  author = {Mo, Man Yue},
  year = {2008},
  journal = {Int. Math. Res. Notices},
  volume = {2008},
  pages = {rnn042},
  doi = {10.1093/imrn/rnn042},
  url = {https://doi.org/10.1093/imrn/rnn042},
}

@article{Parez2022,
	title = {Symmetry-resolved Rényi fidelities and quantum phase transitions},
	volume = {106},
	doi = {110.1103/PhysRevB.106.235101},
	journal = {Phys. Rev. B},
	author = {Parez, Gilles},
	pages = {235101}, 
	year = {2022},
}

@article{Parez2026,
doi = {10.1088/1742-5468/ae3380},
url = {https://doi.org/10.1088/1742-5468/ae3380},
year = {2026},
publisher = {IOP Publishing},
volume = {2026},
number = {1},
pages = {013103},
author = {Parez, Gilles and Alba, Vincenzo},
title = {Reduced fidelities for free fermions out of equilibrium: from dynamical quantum phase transitions to Mpemba effect},
journal = {JSTAT},
}

@article{perez-garcia,
	title = {Dynamical quantum phase transitions from random matrix theory},
	volume = {8},
	doi = {10.22331/q-2024-02-29-1271},
	journal = {Quantum},
	author = {Pérez-García, David and Santilli, Leonardo and Tierz, Miguel},
	year = {2024},
}

@article{silva08,
  title = {Statistics of the Work Done on a Quantum Critical System by Quenching a Control Parameter},
  author = {Silva, Alessandro},
  year = {2008},
  journal = {Phys. Rev. Lett.},
  volume = {101},
  number = {12},
  pages = {120603},
  doi = {10.1103/PhysRevLett.101.120603},
  url = {https://link.aps.org/doi/10.1103/PhysRevLett.101.120603},
}

@article{tierz2025,
       title = {Work Statistics of Sudden Quantum Quenches: {{A}} Random Matrix Theory Perspective on {Gaussianity} and Its Deviations},
      author = {Tierz, Miguel},
      year = {2025},
      journal = {arXiv:2509.09640},
      url = {https://arxiv.org/abs/2509.09640},
      doi = {10.48550/arXiv.2509.09640}
 }

@article{viti2016,
	title = {Inhomogeneous quenches in a free fermionic chain: {Exact} results},
	volume = {115},
	shorttitle = {Inhomogeneous quenches in a free fermionic chain},
	url = {https://iopscience.iop.org/article/10.1209/0295-5075/115/40011},
	doi = {10.1209/0295-5075/115/40011},
	number = {4},
	journal = {EPL},
	author = {Viti, Jacopo and Stéphan, Jean-Marie and Dubail, Jérôme and Haque, Masudul},
	year = {2016},
	pages = {40011},
}

@article{WADIA,
title = {N$=\infty$ phase transition in a class of exactly soluble model lattice gauge theories},
journal = {Phys. Lett. B},
volume = {93},
number = {4},
pages = {403--410},
year = {1980},
doi = {https://doi.org/10.1016/0370-2693(80)90353-6},
url = {https://www.sciencedirect.com/science/article/pii/0370269380903536},
author = {Spenta R. Wadia}
}

@incollection{baik2001,
     author={Baik, J.},
     title={Riemann-{H}ilbert problems for last passage percolation}, 
     booktitle  = {Recent Developments in Integrable Systems and Riemann-Hilbert Problems},
     editor  = {Kenneth D. T.-R. Mclaughlin and Xin Zhou},
     year = {2003},
     pages = {1--22},
     publisher = {American Mathematical Society},
     address = {Providence, R. I.},
}

@book{Abramowitz1964,
author = {Abramowitz, M. and Stegun, I.A.},
title= {Handbook of mathematical functions with formulas, graphs, and mathematical tables},
year = {1964},
address = {New York},
publisher = {Dover},
}

@book{Byrd,
	title = {Handbook of Elliptic Integrals for Engineers and Scientists},
	publisher = {Springer},
	author = {Byrd, P. F. and Friedman, M. D.},
	year = {1971}
}

@book{Bottcher,
	author = {B\"{o}ttcher, A. and Silbermann, B.},
	title = {Analysis of Toeplitz operators},
	publisher = {Springer-Verlag},
	address = {Berlin},
	year = {2006},
	url = {https://doi.org/10.1007/3-540-32436-4},
	doi = {10.1007/3-540-32436-4}
}

@article{copetti2022delayed,
  author        = {Christian Copetti and Alba Grassi and Zohar Komargodski and Luigi Tizzano},
  title         = {Delayed Deconfinement and the Hawking-Page Transition},
  journal       = {JHEP},
  volume        = {04},
  pages         = {132},
  year          = {2022},
  eprint        = {2008.04950},
  archivePrefix = {arXiv},
  primaryClass  = {hep-th}
}

@misc{wolfram_elliptic,
  author       = {{Wolfram Research}},
  title        = {Wolfram Functions: Elliptic Integrals},
  howpublished = {\url{https://functions.wolfram.com/EllipticIntegrals/}},
}

\end{document}